\documentclass[conference]{IEEEtran}

\usepackage{algorithm}
\usepackage{algorithmic}
\usepackage{amsmath}
\usepackage{amsfonts}
\usepackage{amssymb}
\usepackage{array}
\usepackage{cite}
\usepackage{color}
\usepackage{graphicx}
\usepackage{subfigure}
\usepackage{url}

\graphicspath{{./figures/}}
\DeclareGraphicsExtensions{.eps}

\ifCLASSOPTIONcompsoc
  \usepackage[caption=false,font=normalsize,labelfont=sf,textfont=sf]{subfig}
\else
  \usepackage[caption=false,font=footnotesize]{subfig}
\fi

\newtheorem{theorem}{Theorem}

\allowdisplaybreaks[1]
\newcommand\numberthis{\addtocounter{equation}{1}\tag{\theequation}}

\begin{document}

\title{Dynamic Switch-Controller Association and \\Control Devolution for SDN Systems}
\author{\IEEEauthorblockN{Xi Huang$^{1}$, Simeng Bian$^{1}$, Ziyu Shao$^{1}$, Hong Xu$^{2}$}
\IEEEauthorblockA{$^{1}$School of Information Science and Technology, ShanghaiTech University\\
 $^{2}$ NetX Lab @ City University of Hong Kong \\
 Email: \{huangxi,biansm,shaozy\}@shanghaitech.edu.cn, henry.xu@cityu.edu.hk}}\maketitle

\begin{abstract}
In software-defined networking (SDN), as data plane scale expands, scalability and reliability of the control plane have become major concerns. To mitigate such concerns, two kinds of solutions have been proposed separately. One is multi-controller architecture, \textit{i.e.}, a logically centralized control plane with physically distributed controllers. The other is control devolution, \textit{i.e.}, delegating control of some flows back to switches. Most of existing solutions adopt either static switch-controller association or static devolution, which may not adapt well to the traffic variation, leading to high communication costs between switches and controller, and high computation costs of switches.  
In this paper, we propose a novel scheme to jointly consider both solutions, \textit{i.e.}, we dynamically associate switches with controllers and dynamically devolve control of flows to switches. Our scheme is an efficient online algorithm that does not need the statistics of traffic flows. By adjusting a parameter, we can make a trade-off between costs and queue backlogs. Theoretical analysis and extensive simulations show that our scheme yields much lower costs or latency compared to other schemes, as well as balanced loads among controllers. 
\end{abstract}

In the last decade, cloud computing has emerged as the most influential computing paradigm to enable on-demand service hosting and delivery. Despite its importance, efficient resource allocation and network management in data centers are still main challenges to cloud providers. 

Previous works have proposed a variety of solutions to related problems, such as ensemble routing\cite{shao2013intra}, energy budgeting\cite{islam2016online}, workflow scheduling\cite{li2015cost}, virtual slice provisioning\cite{nguyen2015environment}, VM placement\cite{zhao2015online}, etc. 
Meanwhile, software-defined networking (SDN) provides an alternative perspective to manage the whole network. The key idea of SDN is to decouple the control plane from the data plane \cite{mckeown2008openflow}. In such a way, data plane can focus on performing basic functionalities such as packet forwarding at high speed, while the logically centralized control plane manages the whole network. Usually, switches send requests to the control plane for processing some flow events, \textit{e.g.}, flow install events.

The control plane is a potential bottleneck of SDN in terms of scalability and reliability. As the data plane expands, control plane may not be able to process the increasing number of requests if implemented with a single controller, resulting unacceptable latency to flow setup. Reliability is also an issue since a single controller is a single point of failure, which may result in the break-down of the control plane and the entire network. 

Existing proposals to address such problems fall broadly into two categories. One is to implement the control plane as a distributed system with multiple controllers \cite{koponen2010onix}\cite{tootoonchian2010hyperflow}. Each switch then associates with a controller for fault-tolerance and load balancing \cite{levin2012logically}\cite{dixit2013towards}\cite{krishnamurthy2014pratyaastha}\cite{wang2016dynamic}. The other is to devolve part of request processing from controllers to switches to reduce the workload of controllers \cite{curtis2011devoflow}\cite{hassas2012kandoo}\cite{zheng2015lazyctrl}.

For switch-controller association, the first category of solution, the usual design choice is to make a static switch-controller association \cite{koponen2010onix} \cite{tootoonchian2010hyperflow}. However, such static association may result in overloading of controllers and increasing flow setup latency due to its inflexibility to handle traffic variations. An elastic distributed controller architecture is proposed in \cite{dixit2013towards}, with an efficient protocol to migrate switches across controllers. However, it remains open how to determine the switch-controller association. Then Krishnamurthy et al. in \cite{krishnamurthy2014pratyaastha} take a step further by formulating the controller association problem as an integer linear problem with prohibitively high computational complexity. A local search algorithm is proposed to find suboptimal associations within a given time limit (\textit{e.g.}, 30 seconds). In \cite{wang2016dynamic}, the controller is modeled as a M/M/1 queue. Under such an assumption, the controller association problem with a steady-state objective function is formulated as a many-to-one stable matching problem with transfers. Then a novel two-phase algorithm was proposed to connect stable matching with utility-based game theoretic solutions, \textit{i.e.}, coalition formation game with Nash stable solutions.  

For control devolution, the second category of solution, the usual design choice is to statically delegate certain functions and certain flows \cite{curtis2011devoflow} \cite{hassas2012kandoo}\cite{zheng2015lazyctrl}. It remains open how to dynamically delegate in face of traffic variations.  

Based on the above, we identify several interesting questions regarding the control plane design that we try to address:
\begin{itemize}
  \item Instead of deterministic switch-controller association with infrequent re-association \cite{krishnamurthy2014pratyaastha} \cite{wang2016dynamic}, can we directly perform dynamic association with respect to traffic variation? What is the benefit of fine-grained control at the request level?
  \item How to perform dynamic devolution?
  \item How to make a trade-off between dynamic switch-controller association and dynamic control devolution?   
\end{itemize}

In this paper, we consider a general SDN network with traffic variations, incurring dynamic requests to handle flow events. We assume each request can be either processed at a switch (with computation costs) or be uploaded to certain controllers (with communication costs).\footnote{The scenario that some requests can only be processed by a controller is a special case of our model.} We aim at reducing the computational cost by control devolution at data plane, the communication cost by switch-user association between data plane and control plane, and the response time experienced by switches, which is mainly caused by queueing delay on controllers. 

Under such settings, we provide a new perspective and a novel scheme to answer those questions. To the best of our knowledge, this paper is the first to study the joint optimization problem of dynamic switch-controller association and dynamic control devolution. The following are our contributions in this paper.

In the first place, we formulate the problem stated above as a stochastic network optimization problem. Our formulation aims at minimizing the long-term time-average sum of communication cost and computational cost, while keeping time-average queue backlogs of both switches and controllers small. \footnote{By applying \emph{Little's law}, small queue backlog implies small queueing delay or short response time.}

Then, by adopting Lyapunov drift technique \cite{neely2010stochastic} and exploiting sub-problem’s structure, we develop an efficient greedy algorithm to achieve optimality asymptotically. Our algorithm is online, which means it does not need the statistics of traffic workloads and does not need the prior assumption of traffic distribution. In addition to that, our algorithm is also the first to perform the control decisions at the granularity of request level. Note that request-level information such as time-varying queue backlog sizes and number of request arrivals presents the actual time-varying state of data plane. Hence it will help for more accurate decision making of dynamic association and dynamic devolution when compared to coarse-grained control.

Next, we show that our algorithm yields a tunable trade-off between $O(1/V)$ deviation from minimum long-term average sum of communication cost and computational cost and $O(V)$ bound for long-term average queue backlog size. We also find that the positive parameter $V$ determines the switches' willingness of uploading requests to controllers, \textit{i.e.}, performing switch-controller association. We also discuss about two methods to deploy our scheme, along with their advantages and disadvantages in practice.

Last but not least, we conduct large-scale trace-driven simulations to evaluate the performance of our algorithm. Specifically, we run the simulation with four well-known data center networking topologies, viz., Fat-tree topology\cite{al2008scalable}, Canonical 3-Tiered topology\cite{benson2010network}, F10\cite{liu2013f10}, and Jellyfish\cite{singla2012jellyfish}. Simulation results verify the effectiveness and the trade-off of our algorithm. Further, in the extreme case that without control devolution, we compare our dynamic association scheme with other association schemes including Static, Random, and JSQ (Join-the-Shortest-Queue). Simulation results show the advantages of our scheme.
 
We organize the rest of paper as follows. We present the basic idea and formulation in Section 2. Then we show our algorithm design and corresponding performance analysis in Section 3. In Section 4, we present and analyze the simulation results. We conclude this paper in Section 5. 

\begin{figure}[!t]
\centering
 {
 \includegraphics[width=1.0\columnwidth]{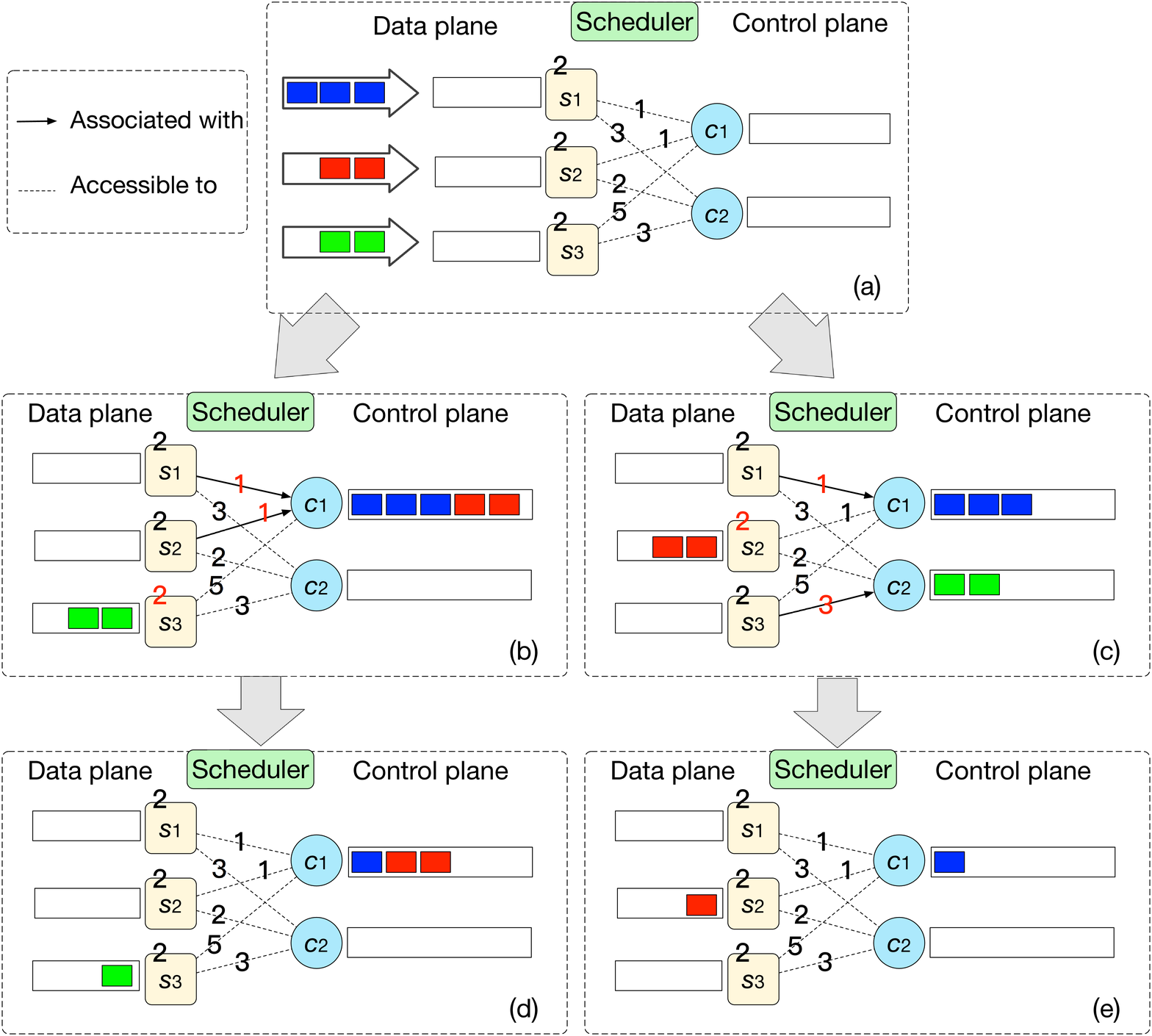}
 }
 \caption{An example that shows the request-level scheduling process. 
There are $3$ switches$(s_1,s_2,s_3)$, $2$ controllers$(c_1,c_2)$, and $1$ global scheduler. The potential connections between controllers and switches are denoted by the dotted lines, while the actual connections (determined by the association) are denoted by the solid lines.
Each switch or controller maintains a queue that buffers requests. 
During each time slot, each switch can decide to store and process its requests locally or to upload requests to some controller. Each controller can serve $2$ requests while each switch can serve only $1$ request. There is a computational cost ($2$ per request on each switch) from local processing by switches themselves, and a communication cost per request if switches upload requests to controllers. For instance, the communication cost is $1$ per request from $s_1$ to $c_1$ and $3$ per request from $s_1$ to $c_2$.  
At the beginning of time slot $t$, $s_1$, $s_2$, and $s_3$ generates $3$, $2$, and $2$ requests, respectively. The scheduler then collects system dynamics and decides a switch-controller association (could be (b) or (c)), aiming at minimizing the sum of communication cost (could be the number of hops, RTTs, etc.) and computational cost, as well as maintaining small queue backlog size. Each switch chooses to either locally process its requests or send them to controllers according to the scheduling decision.}
 \label{fig_sim}
\end{figure}

\section{Problem Formulation}

In this section, we first provide a motivating example for the dynamic switch-controller association and dynamic control devolution. Then we introduce the system model and problem formulation.

\subsection{Motivating Example}

The example of dynamic association and devolution is shown in Fig. \ref{fig_sim}.

First, we focus on the behavior of $s_3$. In Fig. \ref{fig_sim} (b), $s_3$ chooses to process its requests locally,  and that incurs a computational cost of 2 per request. In Fig. \ref{fig_sim} (c), $s_3$ decides to upload requests to $c_2$ and that incurs a communication cost of 3 per request. Although the computational cost is less than communication cost, the decision of locally processing leaves one request not processed yet at the end of the time slot. Hence, it is not necessarily a smart decision for a switch to perform control devolution when its computational cost is lower than its communication cost. Instead, the scheduler should jointly decide control devolution and switch-controller association at the same time.

Next, we focus on the behavior of associations. Fig. \ref{fig_sim} (b) and (c) show two different associations. Fig. \ref{fig_sim} (b) shows the switch-controller association with $(s_1, c_1)$ and $(s_2, c_1)$ ($s_3$ processes requests locally), denoted by $X_1$. In Fig. \ref{fig_sim}, we can see $X_1$ results in uneven queue backlogs, leaving four requests unfinished at the end of the time slot, although it incurs the total cost of communication and computation by only $9$.
Fig. \ref{fig_sim} (c) shows another association with $(s_1, c_1)$ and $(s_3, c_2)$ ($s_2$ processes requests locally), denoted by $X_2$. In Fig. \ref{fig_sim}(e), we can see $X_2$ does better in balancing queue backlogs than $X_1$, but it incurs higher cost by $13$. Thus there is a non-trivial trade-off between minimizing the total cost of communication and computation and maintaining small queue backlogs on each controller.

\subsection{Problem Formulation}

We consider a time slotted network system, indexed by $\{0,1,\,2,\,\dots\}$. Its control plane comprises a set $\mathcal{C}$ of physically distributed controllers, while its data plane consists of a set of switches $\mathcal{S}$. 
Each switch $i \in \mathcal{S}$ keeps a queue backlog of size $Q^s_i(t)$ for locally processing requests, while each controller $j \in \mathcal{C}$ maintains a queue backlog $Q^c_j(t)$ that buffers requests from data plane. We denote $[ Q^c_1(t),\,\dots,\,Q^c_{|\mathcal{C}|}(t) ]$ as $\mathbf{Q}^c(t)$ and $[ Q^s_1(t),\,\dots,\,Q^s_{|\mathcal{S}|}(t) ]$ as $\mathbf{Q}^s(t)$. We use $\mathbf{Q}(t)$ to denote $[\mathbf{Q}^s(t), \mathbf{Q}^c(t)]$.

At the beginning of time slot $t$, each switch $i \in \mathcal{S}$ generates some amounts $0 \le A_i(t) \le a_{max}$ of requests. Then each switch could choose to process its requests either locally or by sending to its associated controller. We assume that each switch $i \in \mathcal{S}$ has a service rate $0 \le U_i(t) \le u_{max}$ to process the devoluted requests, while each controller $j \in \mathcal{C}$ has an available service rate $0 \le B_j(t) \le b_{max}$. We denote $[ A_1(t),\,\dots,\,A_{|\mathcal{S}|}(t) ]$ as $\mathbf{A}(t)$, $[ B_1(t),\,\dots,\,B_{|\mathcal{C}|}(t) ]$ as $\mathbf{B}(t)$, and $\left[ U_1(t),\,\dots,\,U_{|\mathcal{S}|}(t) \right]$ as $\mathbf{U}(t)$. For $i \in \mathcal{S}$ and $j \in \mathcal{C}$, we assume that all $A_i(t)$, $B_j(t)$, and $U_i(t)$ are i.i.d.; besides, $E\{ \left(A_i(t)\right)^2 \} < \infty$, $E\{ \left( B_j(t) \right)^2 \} < \infty$, and $E\{ \left( U_i(t) \right)^2 \} < \infty$.


Then the scheduler collects system dynamics information $\left( \mathbf{A}(t),\,\mathbf{B}(t), \mathbf{Q}(t) \right)$ during current time slot and makes a scheduling decision, denoted by an association matrix $\mathbf{X}(t) \in \{0,\,1\}^{|\mathcal{S}| \times |\mathcal{C}|}$. Here $\mathbf{X}(t)_{i,j} = 1$ if switch $i$ will be associated with controller $j$ during current time slot and $0$ otherwise. An association is feasible if it guarantees that each switch is associated with at most one controller during each time slot. We denote the set of feasible associations as $\mathcal{A}$,
\begin{equation}\label{constraint set}
	\begin{array}{c}
		\displaystyle	\mathcal{A} \triangleq \left\{ \mathbf{X} \in \{0,1\}^{|\mathcal{S}| \times |\mathcal{C}|}\,|\,\sum_{j \in \mathcal{C}} \mathbf{X}_{i,j} \, \le \, 1 \text{ for } i \in \mathcal{S} \right\}
	\end{array}
\end{equation}

According to the scheduling decision, each switch $i$ sends its request to controller $j$ if $\mathbf{X}_{i,j} = 1$. However, if $\sum_{j \in \mathcal{C}} \mathbf{X}_{i,j} = 0$, switch $i$ appends its requests to local queue backlog. Then both switches and controllers serve as many requests in their queues as they could. 
As a result, the update equation for $Q^s_i(t)$ at switch $i$ is 
\begin{equation}\label{ueq for s}
\begin{array}{c}
	\displaystyle Q^s_i(t+1) = \left[ Q^s_i(t) + \left( 1 - \sum_{j \in \mathcal{C}} \mathbf{X}_{i,j}(t) \right) {A}_{i}(t) - U_i(t)\right]^{+}
\end{array}
\end{equation}
and the update equation for $Q^c_j(t)$ at controller $j$ is given by
\begin{equation}\label{ueq for c}
\begin{array}{c}
	\displaystyle Q^c_j(t+1) = \left[ Q^c_j(t) + \sum_{i \in \mathcal{S}} \mathbf{X}_{i,j}(t)\cdot {A}_{i}(t) - B_j(t)\right]^{+}
\end{array}
\end{equation}
where $\left[\,x\,\right]^{+} = \max(x,0)$. 

Having covered the necessary notations and queueing dynamics, we turn to the objective and constraints of our problem. 

\subsubsection{Time-Average Communication Cost}

We define the communication cost between switch $i$ and controller $j$ as $W_{i,j}$ \footnote{The communication cost can be the number of hops or round-trip times (RTT).}. Accordingly, we have a communication cost matrix $\mathbf{W}=\{W_{i,j}\}$. Fixing some association $\mathbf{X} \in \mathcal{A}$, the communication cost within one time slot is
\begin{equation}\label{def-commcost}
	\begin{array}{cl}
		\displaystyle f_{\mathbf{X}}(t) = \hat{f}(\mathbf{X}, \mathbf{A}(t)) \triangleq \sum_{j \in \mathcal{C}} \sum_{i \in \mathcal{S}} W_{i,j} \cdot \mathbf{X}_{i,j} \cdot A_i(t)
	\end{array}
\end{equation}
where we can view $W_{i,j}$ as the price of transmitting one request from switch $i$ to controller $j$. Then, given a series of associations $\{ \mathbf{X}_{0}, \mathbf{X}_{1}, \dots, \mathbf{X}_{t-1} \}$, the time-average expectation of communication cost is shown as follows
\begin{equation}\label{avg-commcost}
	\begin{array}{c}
		\displaystyle \bar{f}(t) \triangleq \frac{1}{t} \sum_{\tau = 0 }^{t - 1} E\left\{f_{\mathbf{X}_{\tau}}(\tau)\right\}
	\end{array}
\end{equation}

\subsubsection{Time-average Computational Cost}

There is a computational cost $\alpha_i$ for each devoluted request to switch $i$ when switch $i$ appends its requests to its local queue backlog for processing. Given some association $\mathbf{X} \in \mathcal{A}$, we define the one-time-slot computational cost as
\begin{equation}\label{def-offload}
	\begin{array}{cl}
		\displaystyle g_{\mathbf{X}}(t) = \hat{g}(\mathbf{X}, \mathbf{A}(t)) \triangleq \sum_{i \in \mathcal{S}} 
		\alpha_i \cdot \left( 1 - \sum_{j \in \mathcal{C}}  \mathbf{X}_{i,j} \right) \cdot A_i(t)
	\end{array}
\end{equation}
Given a series of associations $\{ \mathbf{X}_{0}, \mathbf{X}_{1}, \dots, \mathbf{X}_{t-1} \}$, the time-average expectation of computational cost is
\begin{equation}\label{avg-offload}
	\begin{array}{c}
		\displaystyle \bar{g}(t) \triangleq \frac{1}{t} \sum_{\tau = 0 }^{t - 1} E\left\{g_{\mathbf{X}_{\tau}}(\tau)\right\}
	\end{array}
\end{equation}

\subsubsection{Queueing Stability}

In this paper, we say that a queueing process $\left\{{Q}(t)\right\}$ is stable, if the following condition holds:
\begin{equation}
	\begin{array}{c}
		\displaystyle \lim_{t \to \infty} \frac{1}{t} \sum_{\tau=0}^{t-1} E\left\{Q(\tau)\right\} < \infty \\ 
	\end{array}
\end{equation}

Accordingly, on the data plane, the queueing process $\left\{ \mathbf{Q}^s(t) \right\}$ is stable if
\begin{equation}\label{stabilty1}
	\begin{array}{c}
		\displaystyle \lim_{t \to \infty} \frac{1}{t} \sum_{\tau=0}^{t-1} \sum_{i \in \mathcal{S}} E\left\{Q^s_i(\tau)\right\} < \infty
	\end{array}
\end{equation}

Likewise, on the control plane, the queueing process $\left\{ \mathbf{Q}^c(t) \right\}$ is stable if
\begin{equation}\label{stabilty2}
	\begin{array}{c}
		\displaystyle \lim_{t \to \infty} \frac{1}{t} \sum_{\tau=0}^{t-1} \sum_{j \in \mathcal{C}} E\left\{Q^c_j(\tau)\right\} < \infty
	\end{array}
\end{equation}

Queueing stability implies that both switches and controllers would process buffered requests timely, so that queueing delay is controlled within a limited range.

Consequently, our problem formulation is given as follows 
	\begin{equation}\label{ori-opt-pr}
		\begin{array}{cl}
			\displaystyle \underset{
			\mathbf{X}(t) \in \mathcal{A} \text{ for } t \in \{ 0,1,2,\dots \}
			}{\text{Minimize}} & \displaystyle \lim_{t \to \infty} \sup \left( \bar{f}(t) + \bar{g}(t) \right) \\
			\displaystyle \text{subject to} &  (\ref{ueq for s}), (\ref{ueq for c}), (\ref{stabilty1}), (\ref{stabilty2}).  \\
		\end{array}
	\end{equation}

\section{Algorithm Design and Performance Analysis} 

In this section, we solve our stochastic optimization problem (\ref{ori-opt-pr}) by first transforming it into a series of one-time-slot problems, and designing optimal algorithm that solves the problem in each time slot. Our algorithm design is then followed by a theoretical analysis of its performance.

\subsection{Algorithm Design}

To design a scheduling algorithm that solves problem (\ref{ori-opt-pr}), we adopt the Lyapunov optimization technique in \cite{neely2010stochastic}.

We define the quadratic Lyapunov function as
\begin{equation}\label{lyqueue}
	\begin{array}{cl}
		\displaystyle L(\mathbf{Q}(t)) \triangleq \frac{1}{2} \left( \sum_{j \in \mathcal{C}} \left(Q^c_j(t)\right)^2 + \sum_{i \in \mathcal{S}} \left(Q^s_i(t)\right)^2 \right) \\
	\end{array}
\end{equation}

Next, we define the conditional Lyapunov drift for two consecutive time slots as
\begin{equation}\label{cond-drift}
	\begin{array}{c}
		\Delta\left( \mathbf{Q}(t) \right) \triangleq E\left\{ L(\mathbf{Q}(t+1)) - L(\mathbf{Q}(t))\,|\,\mathbf{Q}(t)\right\}
	\end{array}
\end{equation}

This conditional difference measures the general change in queues' congestion state. We want to push such difference as low as possible, so as to prevent queues $\mathbf{Q}^s(t)$ and $\mathbf{Q}^c(t)$ from being overloaded. However, to maintain small queue backlogs, the action we take, \textit{e.g.} $\mathbf{X}$, might incur considerable communication cost $f_{\mathbf{X}}(t)$ or computational cost $g_{\mathbf{X}}(t)$, or both. Hence, we should jointly consider both queueing stability and the total cost $f_{\mathbf{X}}(t) + g_{\mathbf{X}}(t)$.

Given any feasible association $\mathbf{X} \in \mathcal{A}$, we define the one-time-slot conditional drift-plus-penalty function as
\begin{equation}\label{cond-v-drift}
	\begin{array}{c}
		\Delta_V(\mathbf{Q}(t)) \triangleq \Delta(\mathbf{Q}(t)) + V \cdot E\left\{f_{\mathbf{X}}(t) + g_{\mathbf{X}}(t)|\mathbf{Q}(t)\right\}
	\end{array}
\end{equation}
where $f_{\mathbf{X}}(t)$ is defined by (\ref{def-commcost}), $g_{\mathbf{X}}(t)$ is defined by (\ref{def-offload}), and $V > 0$ is a constant that weights the penalty brought by $f_{\mathbf{X}}(t)$ and $g_{\mathbf{X}}(t)$.

By minimizing the upper bound of the drift-plus-penalty expression (\ref{cond-v-drift}), the time-average communication cost can be minimized while stabilizing the network of request queues\cite{neely2010stochastic}. We then employ the concept of \emph{opportunistically minimizing an expectation} in \cite{neely2010stochastic}, and we transform the long-term stochastic optimization problem (\ref{ori-opt-pr}) into the following drift-plus-penalty minimization problem at every time slot $t$. The details have been relegated to Appendix-A. 
\begin{equation}\label{obj-fn}
	\begin{array}{cl}
	\underset{\mathbf{X} \in \mathcal{A}}{\text{Minimize}} & \displaystyle
	V \cdot \left( \hat{f}(\mathbf{X}, \mathbf{A}(t)) + \hat{g}(\mathbf{X}, \mathbf{A}(t)) \right) + \\
	& \displaystyle \sum_{j \in \mathcal{C}} Q^c_j(t) \cdot \left[ \sum_{i \in \mathcal{S}} \mathbf{X}_{i,j} \cdot A_i(t) \right] + \\
	& \displaystyle \sum_{i \in \mathcal{S}} Q^s_i(t) \cdot \left[ (1 - \sum_{j \in \mathcal{C}} \mathbf{X}_{i,j}) \cdot A_i(t) \right] \\
	\end{array}
\end{equation}

After rearranging the terms in (\ref{obj-fn}), our optimization problem turns out to be

\begin{equation}\label{opt-pr}
	\begin{array}{ll}
		\underset{\mathbf{X} \in \mathcal{A}}{\text{Minimize}} & \displaystyle \sum_{i \in \mathcal{S}} \displaystyle \left[ V \alpha_i + Q^s_i(t) \right] A_i(t) + \sum_{i \in \mathcal{S}} \sum_{j \in \mathcal{C}} 
			\left[ V W_{i,j} + \right. \\
		& \displaystyle 
		\left. Q_j^c(t) - V\alpha_i - Q^s_i(t) \right] \mathbf{X}_{i,j} A_i(t) \\
	\end{array}
\end{equation}

Since the first summing term $\sum_{i \in \mathcal{S}} [V\alpha_i + Q_i^s(t)]$ in (\ref{opt-pr}) has nothing to do with $\mathbf{X}$, then we regard it as constant and focus on minimizing the second term of (\ref{opt-pr}) only. 

For each $i \in \mathcal{S}$, we split $\mathcal{C}$ into two disjoint sets $\mathcal{J}^i_1$ and $\mathcal{J}^i_2$, \textit{i.e.} $\mathcal{J}^i_1 \mathbin{\dot{\bigcup}} \mathcal{J}^i_2 = \mathcal{C}$, and
\begin{equation}\label{ji1}
	\begin{array}{c}
		\mathcal{J}^i_1 \triangleq \{ j \in \mathcal{C} \,|\, V W_{i,j} +  Q_j^c(t) > V\alpha_i + Q^s_i(t) \}
	\end{array}
\end{equation} 
\begin{equation}\label{ji2}
	\begin{array}{c}
		\mathcal{J}^i_2 \triangleq \{ j \in \mathcal{C} \,|\, V W_{i,j} +  Q_j^c(t) \le V\alpha_i + Q^s_i(t) \}		
	\end{array}
\end{equation} 

Then, for each switch $i \in \mathcal{S}$,
\begin{equation}\label{1stterm}
	\begin{array}{cl}
		& \displaystyle 
			\sum_{j \in \mathcal{C}} 
			\left[ V W_{i,j} +  Q_j^c(t) - V\alpha_i - Q^s_i(t) \right] \mathbf{X}_{i,j} A_i(t) \\
		= & \displaystyle 
			\left\{ 
			\sum_{j \in \mathcal{J}^{i}_1} \left[ V W_{i,j} +  Q_j^c(t) - V\alpha_i - Q^s_i(t) \right] \mathbf{X}_{i,j} + \right. \\
		& \displaystyle \left. \sum_{j\in\mathcal{J}^{i}_2} \left[ V W_{i,j} +  Q_j^c(t) - V\alpha_i - Q^s_i(t) \right] \mathbf{X}_{i,j}	
		\right\} A_i(t)
	\end{array}
\end{equation}

Next, we show how to minimize (\ref{1stterm}) with $\mathbf{X} \in \mathcal{A}$. Given any $(i,j) \in \mathcal{S} \times \mathcal{C}$, we define
\begin{equation}\label{weight-fn}
	\begin{array}{cl}
		\omega(i,j) & \triangleq V W_{i,j} +  Q_j^c(t) - V\alpha_i - Q^s_i(t) \\
		& = V \cdot \left( W_{i, j} - \alpha_i \right) + \left( Q_j^c(t) - Q_i^s(t) \right) \\
	\end{array}
\end{equation}

Here, we define $\mathbf{X}^{*}$ as the optimal solution to minimize (\ref{1stterm}). For each switch $i \in \mathcal{S}$, we should consider two different cases.

\begin{itemize}
	\item[i.] If $\mathcal{J}^i_2 = \emptyset$, \textit{i.e.}, $\omega(i,j) > 0$ for all $j \in \mathcal{C}$, then the only way to minimize (\ref{1stterm}) is setting $\mathbf{X}^*_{i,j} = 0$ for all $j \in \mathcal{C}$. 
	\item[ii.] If $\mathcal{J}^i_2 \neq \emptyset$, then we handle with $\mathbf{X}_{i,j}$ for $j \in \mathcal{J}^i_1$ and $j \in \mathcal{J}^i_2$ separately.
	\begin{itemize}
		\item For $j \in \mathcal{J}^i_1$, to minimize (\ref{1stterm}), it is not hard to see we should set $\mathbf{X}^*_{i,j} = 0$ for all $j \in \mathcal{J}^i_1$.\\
		\item For $j \in \mathcal{J}^i_2$, $\omega(i,j) \le 0$.  Then we should make $\mathbf{X}^*_{i,j^*} = 1$ for such $j^*$ that
		\begin{equation}
			\begin{array}{cl}
				j^* = \underset{j \in \mathcal{J}^i_2}{\arg \min} & \omega(i,j)
			\end{array}
		\end{equation} 
	 and $\mathbf{X}^*_{i,j} = 0$ for $j \in \mathcal{J}^i_2-\{j^*\}$. 
	\end{itemize}
\end{itemize}
	 
In such a way, given any $\mathbf{X}' \in \mathcal{A}$, for switch $i$ the following inequality always holds
		\begin{equation}
			\begin{array}{cl}
				& \displaystyle \sum_{j\in\mathcal{J}^{i}_1} \omega(i,j) \cdot \mathbf{X}'_{i,j} + 
				\displaystyle \sum_{j\in\mathcal{J}^{i}_2} \omega(i,j) \cdot \mathbf{X}'_{i,j} \\ 
			    \ge & \displaystyle \left[ \sum_{j\in\mathcal{J}^{i}_1} \omega(i,j) \right] \cdot 0 + \underset{j \in \mathcal{J}^i_2}{\min} \omega(i,j) \\
				= & \displaystyle \sum_{j\in\mathcal{J}^{i}_1} \omega(i,j) \cdot \mathbf{X}^*_{i,j} + 
				\sum_{j\in\mathcal{J}^{i}_2} \omega(i,j) \cdot \mathbf{X}^*_{i,j}
			\end{array}
		\end{equation}

Therefore, the association $\mathbf{X}^{*}$ produced by the above process is the optimal solution that minimizes (\ref{1stterm}), and equivalently (\ref{opt-pr}). 

As a result, we have the algorithm shown as follows:
\begin{algorithm}
 \caption{Greedy Scheduling Algorithm}
 \begin{algorithmic}[1]
 \renewcommand{\algorithmicrequire}{\textbf{Input:}}
 \renewcommand{\algorithmicensure}{\textbf{Output:}}
 \REQUIRE During time slot $t$, the scheduler collects queue lengths information from individual controllers and switches, \textit{i.e.} $\mathbf{Q}^c(t)$, $\mathbf{Q}^s(t)$, and $\mathbf{A}(t)$.
 \ENSURE  A scheduling association $\mathcal{X} \subset \mathcal{S} \times \mathcal{C}$
 \\
  \STATE Start with an empty set $\mathcal{X} \leftarrow \emptyset$ 
  \FOR{each switch $i \in \mathcal{S}$}
  	\STATE Split all controllers $\mathcal{C}$ into two sets $\mathcal{J}^i_1$ and $\mathcal{J}^i_2$, where
  	$\mathcal{J}^i_1 = \{ j \in \mathcal{C} \,|\, \omega(i,j) > 0 \}$ and \\
    $\mathcal{J}^i_2 = \{ j \in \mathcal{C} \,|\, \omega(i,j) \le 0 \}$ \\
    \STATE If $\mathcal{J}^i_2=\emptyset$, then skip current iteration.
    \STATE If $\mathcal{J}^i_2\neq\emptyset$, then choose controller $j^* \in \mathcal{J}^i_2$ such that
		$$\displaystyle j^* \in \underset{j \in \mathcal{C}}{\arg\min}\,\, \omega(i, j)$$
	\STATE $\mathcal{X} \leftarrow \mathcal{X}\bigcup\{(i,j^*)\}$
 \ENDFOR \\
 \RETURN $\mathcal{X}$ \\
 \textit{According to $\mathcal{X}$, switches upload requests to controllers or append requests to their local queues. Then controllers and switches update their queue backlogs as in (\ref{ueq for s}) and (\ref{ueq for c}) after serving requests.} 
 \end{algorithmic}
 \end{algorithm}

\textbf{Remarks:} 
\begin{enumerate}
  \item[i.] Our algorithm is greedy. It is because that
  for each switch $i \in \mathcal{S}$, switch $i$ will upload requests onto control plane, if there exists any controller $j$ such that $\omega(i,j) \le 0$. For the chosen controller $j^*$, by the definition of $\omega(i,j^*)$ in (\ref{weight-fn}), $\omega(i, j^*) < 0$ implies that either $W_{i,j^*} < \alpha_i$ or $Q^c_{j^*}(t) < Q^s_i(t)$. By contrast, switch $i$ will process its requests locally if $\omega(i,j) > 0$ for all $j \in \mathcal{C}$. In other words, our algorithm greedily associates each switch with controllers that either with relatively small queue backlog size or with low communication cost (smaller than the switch's computational cost), and otherwise it leaves all requests locally processed. 
  \item[ii.] For switch $i$, given any controller $j$ such that $W_{i,j} > \alpha_i$, switch $i$ decides to upload requests to $j$ only if $\omega(i,j)$ is non-positive and smaller than any other. This requires switch $i$ itself holds enough requests locally, \textit{i.e.}, $Q^s_i(t) \ge V \cdot \left( W_{i,j} - \alpha_i \right) + Q^c_j(t)$. Then it will upload requests. Thus smaller $V$ will invoke more effectively the willingness of switch $i$ to upload requests to control plane.
  \item[iii.] On the other hand, for switch $i$, given any controller $j$ such that $W_{i,j} < \alpha_i$, switch $i$ will process requests locally if control plane holds large amounts of requests, \textit{i.e.}, $Q^s_i(t) < V \cdot \left( W_{i,j} - \alpha_i \right) + Q^c_j(t)$. Thus given very large $V$, controllers will have to hold great loads of requests before switches become willing to process requests locally.
  \item[iv.] Therefore, the parameter $V$ actually controls switches' willingness of uploading requests to controllers, \textit{i.e.}, performing switch-controller association. In other words, it controls the trade-off between communication cost and the computational cost, which are incurred by uploading requests to control plane and locally processing, respectively. 
\end{enumerate}

\subsection{Performance Analysis}

Next we characterize the performance of our algorithm. We suppose $g^*$ and $f^*$ are the supremum of time-average computational cost and communication cost that we want to achieve, respectively. We also suppose $d_{\max}=\max_{i,j} \left\{ E(B_j^2(t)), E(U_i^2(t)), E(A_i^2(t)) \right\}$. The we have the following theorem on the $O(1/V),O(V)$ trade-off between costs and queue backlogs:
\begin{theorem}
	Given the parameters $V>0$, $\epsilon>0$, and constant $\displaystyle K \ge \frac{d_{\max} \cdot (|\mathcal{C}| + |\mathcal{S}| + |\mathcal{S}|^2)}{2}$, then the queueing vector process $\mathbf{Q}(t)$ is stable; besides, the time-average expectation of communication cost and computational cost, as well as queue backlogs on switches and controllers satisfy:
	\begin{equation}\label{th1}
		\begin{array}{cl}
			i. & \displaystyle \underset{t \to \infty}{\lim \sup} \left( \bar{f}(t) + \bar{g}(t) \right) \le f^* + g^* + \frac{K}{V} \\
			ii. & \displaystyle \underset{t \to \infty}{\lim \sup} \frac{1}{t} \sum_{\tau=0}^{t-1} \left[ \sum_{j \in \mathcal{C}} E\left\{ Q^c_j(\tau) \right\} + \sum_{i \in \mathcal{S}} E\left\{ Q^s_i(\tau) \right\} \right]
			\\
			& \displaystyle \le \frac{K + V \cdot (f^* + g^*) }{\epsilon} \\
	\end{array}
	\end{equation}
\end{theorem}
The proof of theorem 1 is relegated to Appendix-B.  

\section{Simulation Results}

\subsection{Basic Settings}

\begin{figure}[!t]
\centering
 \includegraphics[width=0.45\textwidth]{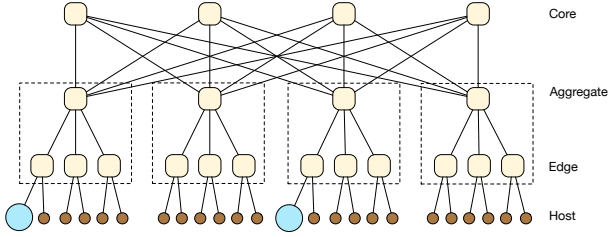}
 \caption{An instance of Canonical 3-Tiered topology with $k=4$, where $k$ denotes the switch port number. In this paper, the number of aggregate switches is also set to $k$, and each connects to $k-1$ edge switches. The total number of switches is $k^2 + k$. Each edge switch is directly connected to $\frac{k}{2}$ hosts. Therefore, there are $\frac{k^3 - k^2}{2}$ hosts in total.} 
 \label{topo_3tiered}
\end{figure}

\begin{figure}[!t]
\centering
 \includegraphics[width=0.45\textwidth]{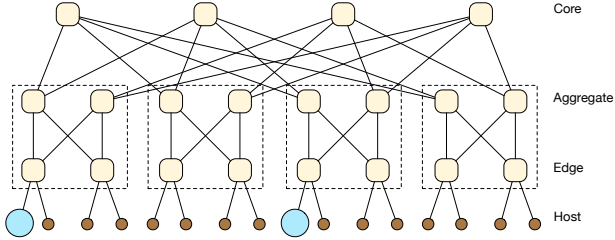}
 \caption{An instance of Fat-tree topology with $k=4$, where $k$ denotes the number of switch ports. The number of core, aggregate, edge switches are $\frac{k^2}{4}$, $\frac{k^2}{2}$, $\frac{k^2}{2}$, respectively. And the total number of switches is $\frac54 k^2$. Each edge switch is directly connected to $\frac{k}{2}$ hosts. Accordingly, there are $\frac{k^3}{4}$ hosts in total.}
\label{topo_fattree}
\end{figure}

\begin{figure}[!t]
\centering
 \includegraphics[width=0.24\textwidth]{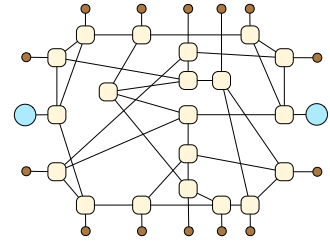}
 \caption{An instance of Jellyfish topology with $k=4$, where $k$ denotes the number of switch ports. Jellyfish topology is created by constructing a random graph at the top-of-rack (ToR) switch layer. It comprises $\frac{5}{4}k^2$ switches and $\frac{k^3}{4}$ hosts, which is comparable to other topologies.
 }
\label{topo_jelly}
\end{figure}

\begin{figure}[!t]
\centering
 \includegraphics[width=0.45\textwidth]{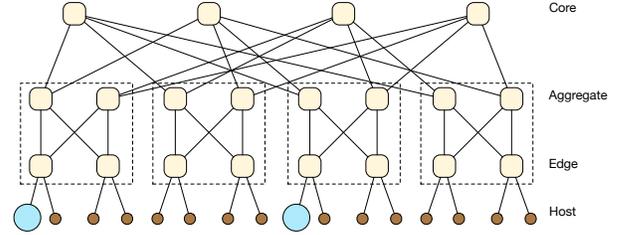}
 \caption{An instance of F10 topology with $k=4$, where $k$ denotes the number of switch ports. Both F10 and Fat-tree with the same port number ($k=4$ in this case) have identical number of switches in each layer (core, aggregate, and edge), as well as identical number of hosts. The only difference lies in how switches connect to their upper-layer switches. F10 breaks the symmetry of Fat-tree's structure by employing AB-tree to enhance its fault tolerance\cite{liu2013f10}.}
\label{topo_f10}
\end{figure}

\textbf{Topology:} We evaluate our {\bf Greedy} scheduling algorithm under four well-known data center topologies: Canonical 3-Tiered topology\cite{benson2010network}, Fat-tree\cite{al2008scalable}, Jellyfish\cite{singla2012jellyfish}, as well as F10\cite{liu2013f10}. We show one instance for each of them, respectively, in Fig. \ref{topo_3tiered} - Fig. \ref{topo_f10}.

To make our performance analysis comparable among the four topologies, we construct instances of these topologies at almost the same scale. In addition, we assume that all switches are identical with the same port number.

Regarding Fat-tree, F10, and Jellyfish topology, we set the switch's port number as $24$. Hence all of them comprise $720$ switches. Specifically, in Jellyfish, switches are wired randomly and each switch connects to $4\sim{5}$ hosts.
Regarding the Canonical 3-Tiered topology, remind that its number of switches is $k^2+k$ in our setting, which is determined by the switch port number $k$. To make it at the same scale as other topologies, we set the switch port number as $26$ and thus there are $702$ switches in total.
Note that these resulting topologies are also comparable to the size of commercial data centers \cite{benson2010network}. 

In these topologies, we deploy controllers on the hosts, which are denoted by the blue circles in Fig. \ref{topo_3tiered}, Fig. \ref{topo_fattree}, Fig. \ref{topo_jelly}, and Fig. \ref{topo_f10}. In deterministic topologies (Fat-tree, Canonical 3-Tiered, and F10), we deploy one controller for every two pods\footnote{In Canonical 3-Tiered topology, we regard the group of switches that affiliate the same aggregation switch as one pod (including the aggregation switch itself).}. In random topology (Jellyfish), we keep the number of controllers the same as in other topologies, and deploy controllers on hosts with non-neighboring ToRs. 

\textbf{Traffic Workloads:} We conduct trace-driven simulations, where the flow arrival process on each switch follows the distribution of flow inter-arrival time in \cite{benson2010network}, which is drawn from measurements within real-world data centers. In \cite{benson2010network}, the average flow inter-arrival time is about $1700 \mu s$. Note that in our simulation, we differentiate flows neither by their sizes, \textit{i.e.}, mice flows and elephant flows, nor by their deadlines, \textit{i.e.}, delay-sensitive flows and the insensitive ones. Nevertheless, our solution leaves the freedom to classify flows and prioritize the requests according to their characteristics. 

 We then set the length of each time slot as $10ms$. Accordingly, the average flow arrival rate on each switch is about $5.88$ flows per time slot. 

In fact, there do exist hot spots within pods in real-world data center networks, where the switches have significantly high flow arrival rates. In our simulation, we pick the first pod as a hot spot and all switches there have significantly high flow arrival rate, \textit{i.e.}, $200$ flows per time slot. As for controllers, we set their individual capacity as $600$ flows per time slot. That is consistent with the capacity of a typical NOX controller \cite{tootoonchian2012controller}.

\textbf{Costs:} Given any network topology, we define the communication cost $W_{i,j}$ between switch $i$ and controller $j$ as the length (number of hops) of shortest path from $i$ to $j$. Then we set a common computation cost $\alpha$ for all switches, which equals to the average hop number between switches and controllers of its underlying topology. In both Fat-tree and F10 topologies, $\alpha = 4.13$; while in 3-Tiered and Jellyfish topologies, $\alpha$ is $4.81$ and $3.56$ (in Jellyfish, it depends on the generated instance), respectively.


\begin{figure}[!t]
\centering
 \subfigure[Centralized] {
 \includegraphics[width=0.45\columnwidth]{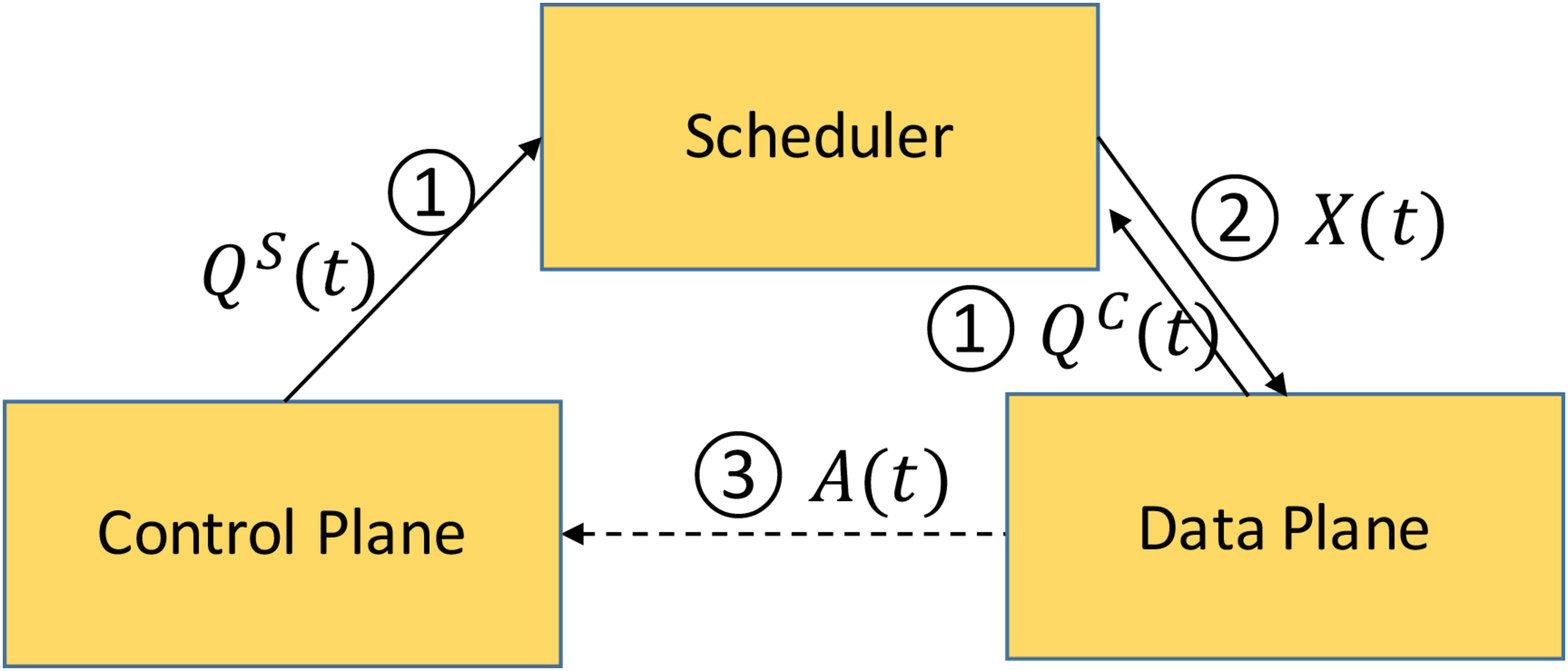}
 }
 \subfigure[Decentralized] {
 \includegraphics[width=0.42\columnwidth]{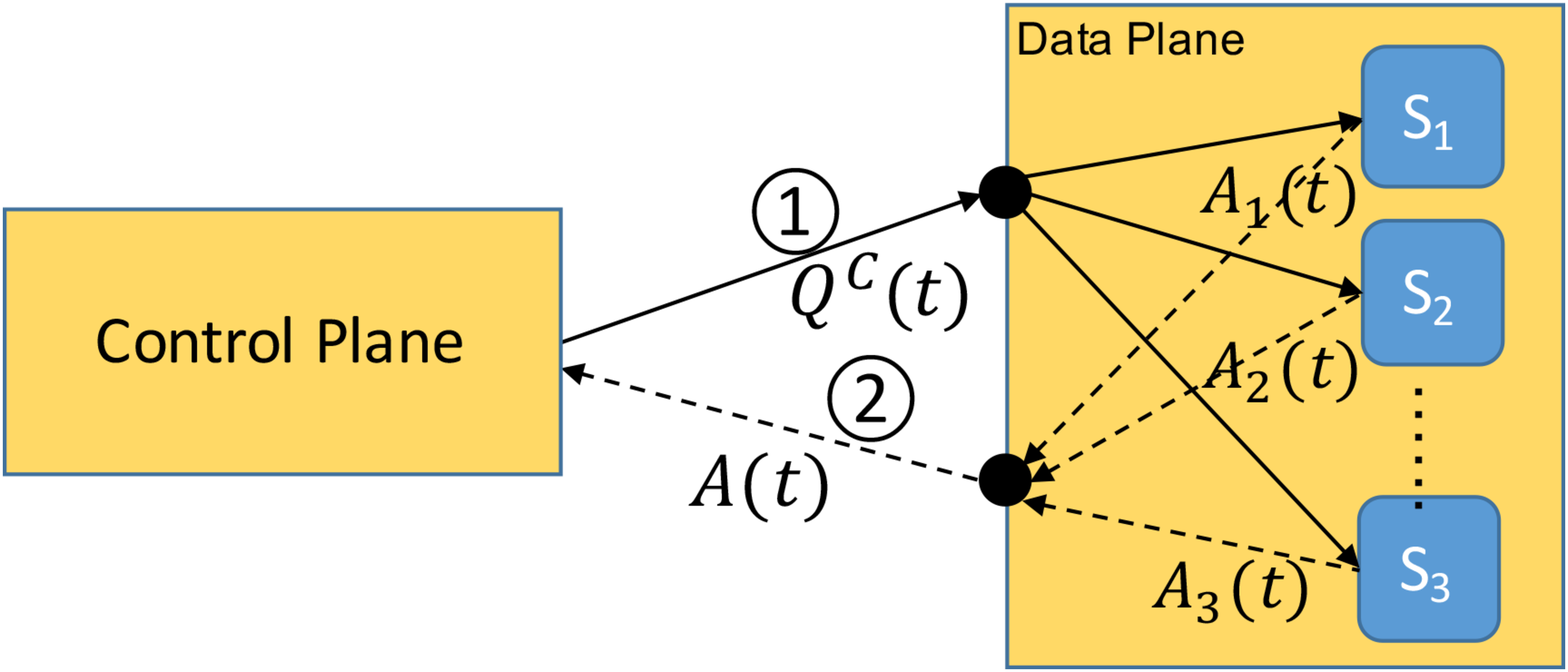}
 }
 \caption{Two scheduler implementations to apply \textbf{Greedy}}
 \label{architecture}
\end{figure}

\textbf{Scheduler implementations:} As Fig. \ref{architecture} shows, our algorithm can be implemented in an either decentralized or centralized manner. 

In the centralized way, the scheduler is independent of both control plane and data plane. The scheduler collects system dynamics including queue backlogs on both switches and controllers to make a centralized scheduling decision. Next, it spreads the scheduling decision onto switches; then switches upload or locally process their requests according to the decision. The abstract process is presented in Fig. \ref{architecture} (a). The advantage of centralized architecture is that it doesn't require modification on data plane, \textit{i.e.}, all the system dynamics such as the communication cost and queue backlogs can be obtained via standard OpenFlow APIs. This is well-suited for the situation where the data plane is at a large scale and switches' compute resource is scarce. In fact, the scheduler could also be deployed on control plane. There are disadvantages, too. Centralized scheduler is a potential single point of failure, or even a bottleneck with considerable computation. Besides, it requires back-and-forth message exchange between the SDN system and the scheduler, which leads to longer response time.  

In the decentralized way, as Fig. \ref{architecture} (b) shows, switches will periodically update their information about queue backlogs in control plane. Then each of them makes independent scheduling decision and processes the requests either locally or on control plane. Though requiring modification on switches, the decentralized way still has the following advantages. It requires less amounts of message exchange than that in the centralized way, thus switches would response even faster to handling flow events. Meanwhile, the computation of our \textbf{Greedy} is distributed onto switches, leading to better scalability and fault tolerance.

\subsection{Evaluation of Greedy Algorithm}

\begin{figure}[!t]
\centering
 \subfigure[Total cost vs. V] {
 \includegraphics[width=0.465\columnwidth]{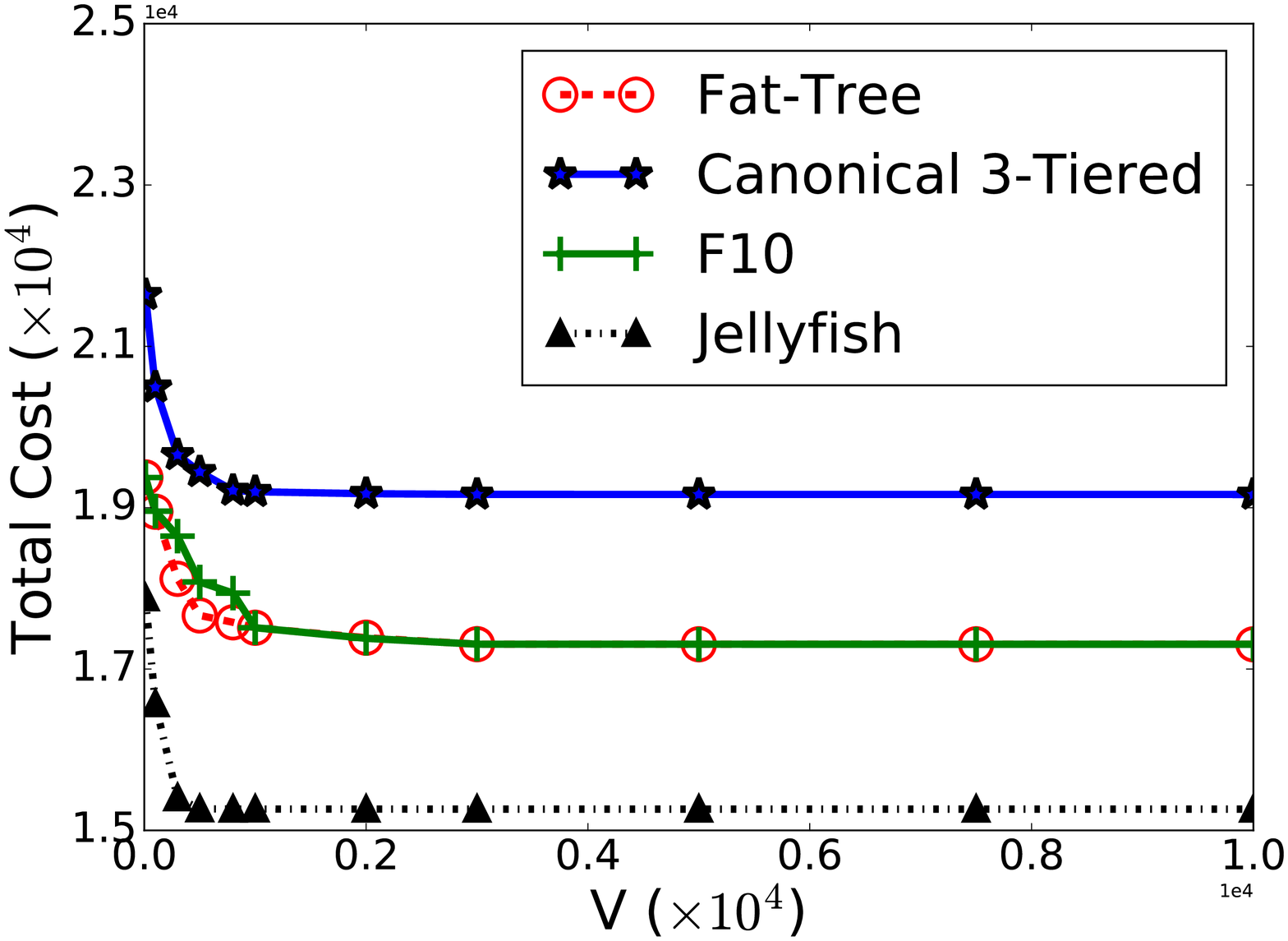}
 }
 \subfigure[Total queue backlog vs. V] {
 \includegraphics[width=0.465\columnwidth]{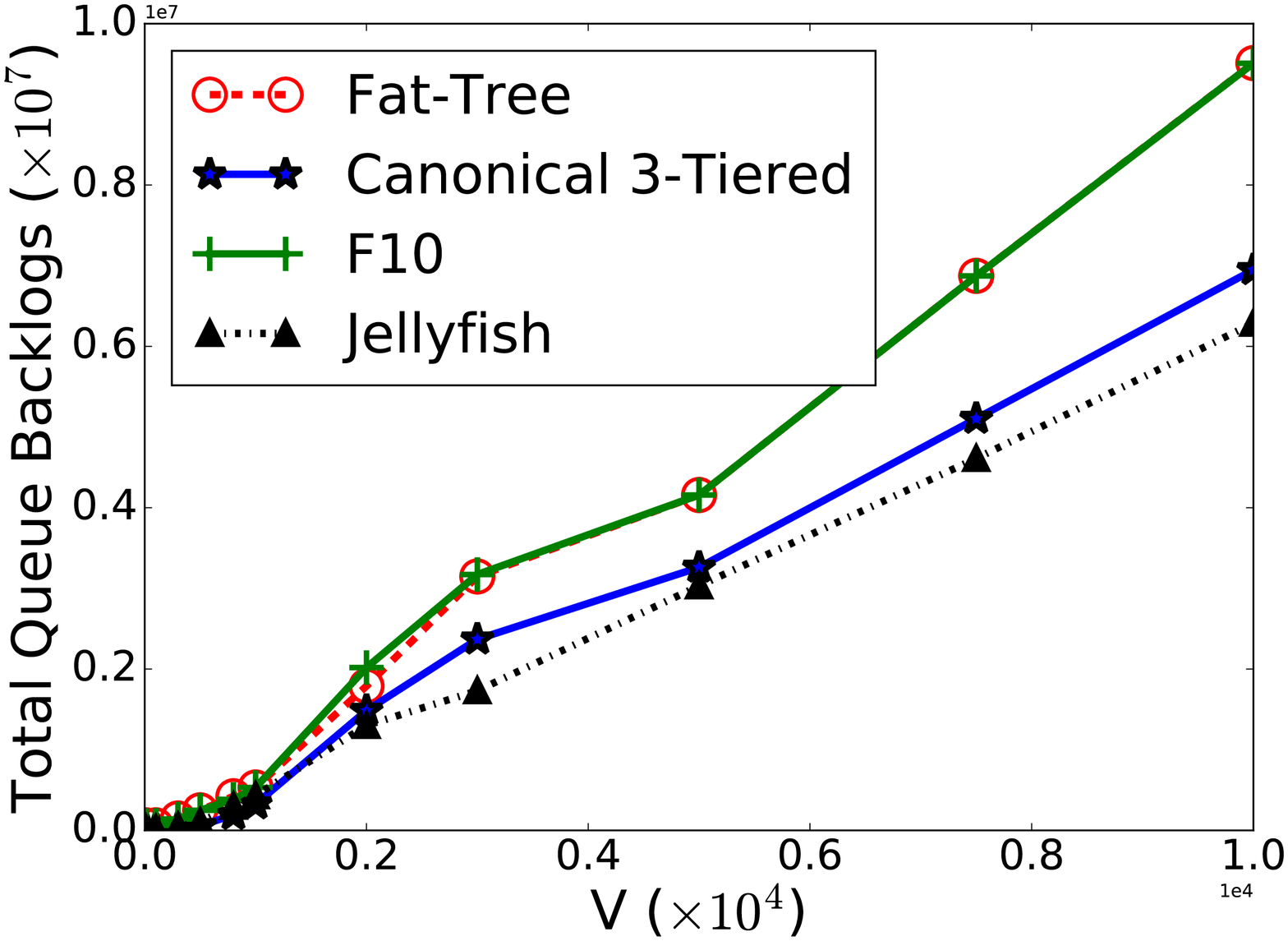}
 }
 \caption{Performance of \textbf{Greedy} under Fat-tree, Canonical 3-Tiered, F10, and Jellyfish topology in terms of (a) the sum of total communication cost and computational cost, and (b) total queue backlog.}
 \label{trace}
\end{figure}

Fig. \ref{trace} (a) presents how the summation of long-term average communication cost and computational cost changes with different $V$ in those four topologies. We make the following observations. 

First, as $V$ varies from $0$ to $1.0 \times 10^4$, it shows that the total cost goes down gradually. This is consistent with our previous theoretic analysis. The intuition behind such decline is as follows. Remind that $V$ controls the switches' willingness of uploading requests. For switches that are close to controllers (their communication cost is less than the average), large $V$ makes them unwilling to process requests locally unless the controllers get too heavy load. As $V$ increases, those switches will choose to upload requests to further reduce the costs since for those switches, communication costs are less than the computation costs. 

Second, the total cost in 3-Tiered topology is more than the other schemes'. The reason is two fold. One is 3-Tiered has a higher computational cost ($\alpha = 4.81$ compared to $4.13$ and $3.56$) and it costs even more when switches process requests locally. The other is when it comes to communication cost, switches in 3-Tiered topology usually take longer path to controllers compared to those in other topologies. 

Third, the total cost in Jellyfish topology is significantly lower than the others. As illustrated in \cite{singla2012jellyfish}, compared to deterministic topologies, Jellyfish takes the advantages that all its paths are on average shorter\footnote{Remind that $\alpha$ is set to be the average path length in our settings.} than in other topologies of the same scale. 

\begin{figure}[!t]
	\centering
	\subfigure[Total cost vs. V] {
		\includegraphics[width=0.45\columnwidth]{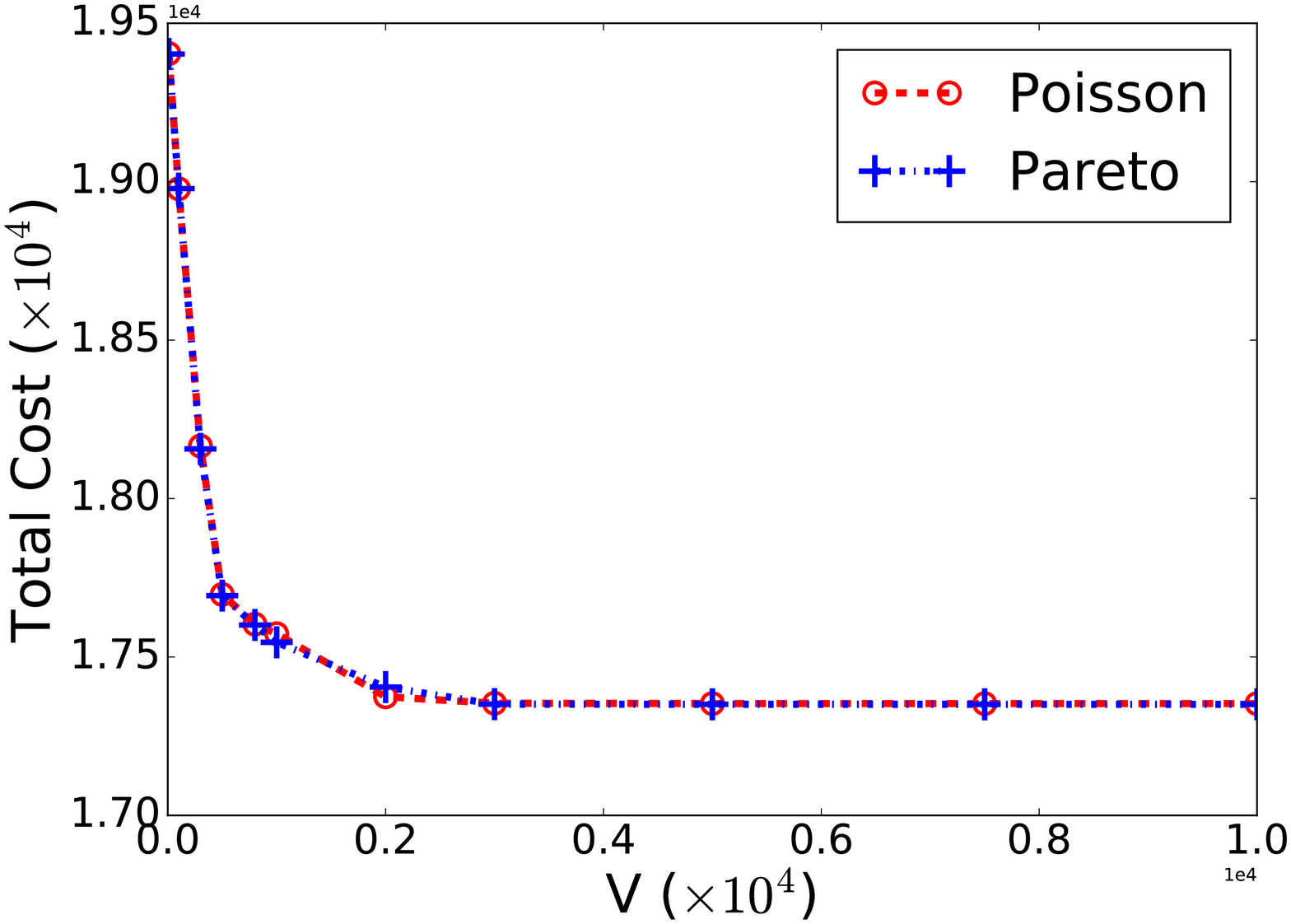}
	}	
	\subfigure[Total queue backlog vs. V] {
		\includegraphics[width=0.45\columnwidth]{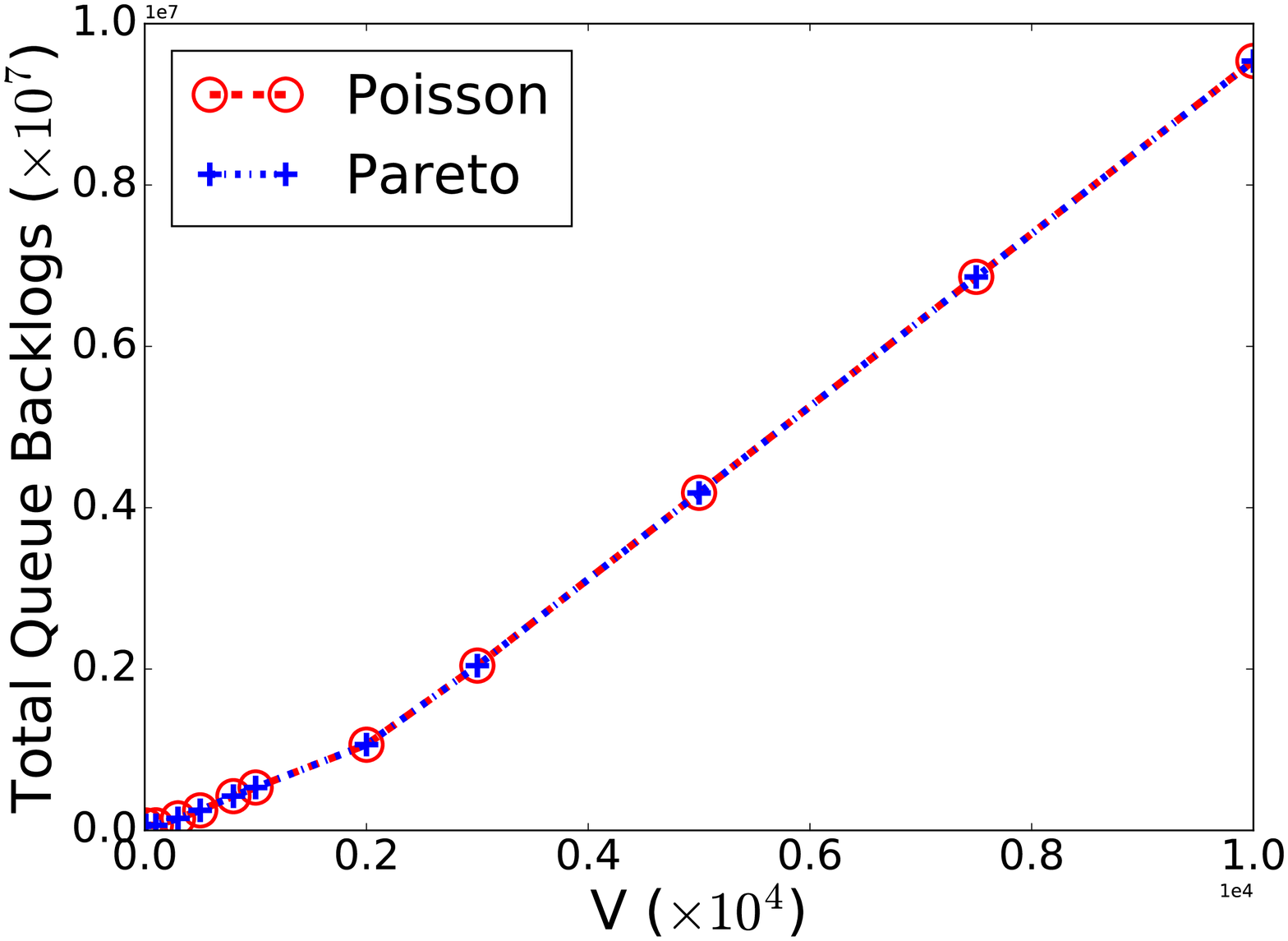}
	}
	
	\caption{Performance of \textbf{Greedy} under Fat-tree topology with request arrivals that follow Poisson and Pareto process in terms of (a) the sum of total communication cost and computational cost, and (b) total queue backlog. For Poisson process, its arrival rate is set to be $5.88$; while for Pareto process, its shape parameter and its scale parameter are set to be $2$ and $2.94$, respectively.}
	\label{pp}
\end{figure}

Fig. \ref{trace} (b) shows the varying of total queue backlog size with different values of $V$. We notice that there is a linear rising trend in total queue backlog size for all four topologies. This is also consistent with the $O(V)$ queue backlog size bound in (\ref{th1}). Recall our analysis in \emph{Total Cost}: larger $V$ invokes most switches to spend more time uploading requests to control plane. However, requests on control plane will keep accumulating since controllers' service capacity is fixed. Thus when $V$ becomes sufficiently large, control plane will eventually hold most of requests in the system. This explains the increasing queue backlog size in Fig. \ref{trace} (b). 

Fig. \ref{pp} (a) shows the total cost of {\bf Greedy} in Fat-tree topology with other two request arrival processes. 
The curves of total cost with Poisson and Pareto almost overlap, with a gradual declined reduction to the minimum. Similarly, in Fig. \ref{pp} (b), we can see the total queue backlog size in Fat-tree topology when we apply {\bf Greedy} with request arrivals that follow Poisson and Pareto processes. The queue backlogs under both arrival processes remain overlapping all the time. Hence Fig. \ref{pp} shows that our scheme doesn't require the statistics of traffic workloads or the prior assumption of traffic distribution.

Note that we do not show the curves here for the other three topologies, because curves are also overlapping as those in Fig. \ref{pp}(a) and Fig. \ref{pp}(b).

\subsection{Comparison with Other Association Schemes}

In this subsection, we consider the extreme case by setting common computational cost $\alpha = 2.0 \times 10^{28}$ for all switches. This means the cost of local processing requests are prohibitively high and at each time slot switches would only choose to upload requests to controllers. Such a setting emulates the scenarios where switches' computing resources are extremely scarce or local processing is not supported. As a result, our greedy algorithm degenerates into a dynamic switch-controller association algorithm. 

We compare \textbf{Greedy}'s performance along with three other schemes: \textbf{Static}, \textbf{Random} and \textbf{JSQ (Join-the-Shorest-Queue)}. In \textbf{Static} scheme, each switch $i$ chooses the controller $j$ with minimum communication cost $W_{i,j} = \min_{k \in \mathcal{C}} W_{i, k}$ and then fixes such an association in all time slots. In \textbf{Random} scheme, each switch is scheduled to pick up a controller uniformly randomly during each time slot. In \textbf{JSQ} scheme, each switch $i$ is scheduled to pick the controller with smallest queue backlogs, among its available candidates. After choosing the target controller, each switch pushes all its available requests (those haven't been put into local processing queue yet) to the controller's queue. 

\begin{figure}[!t]
\centering
 \subfigure[Canonical 3-Tiered topology] {
 \includegraphics[width=0.44\columnwidth]{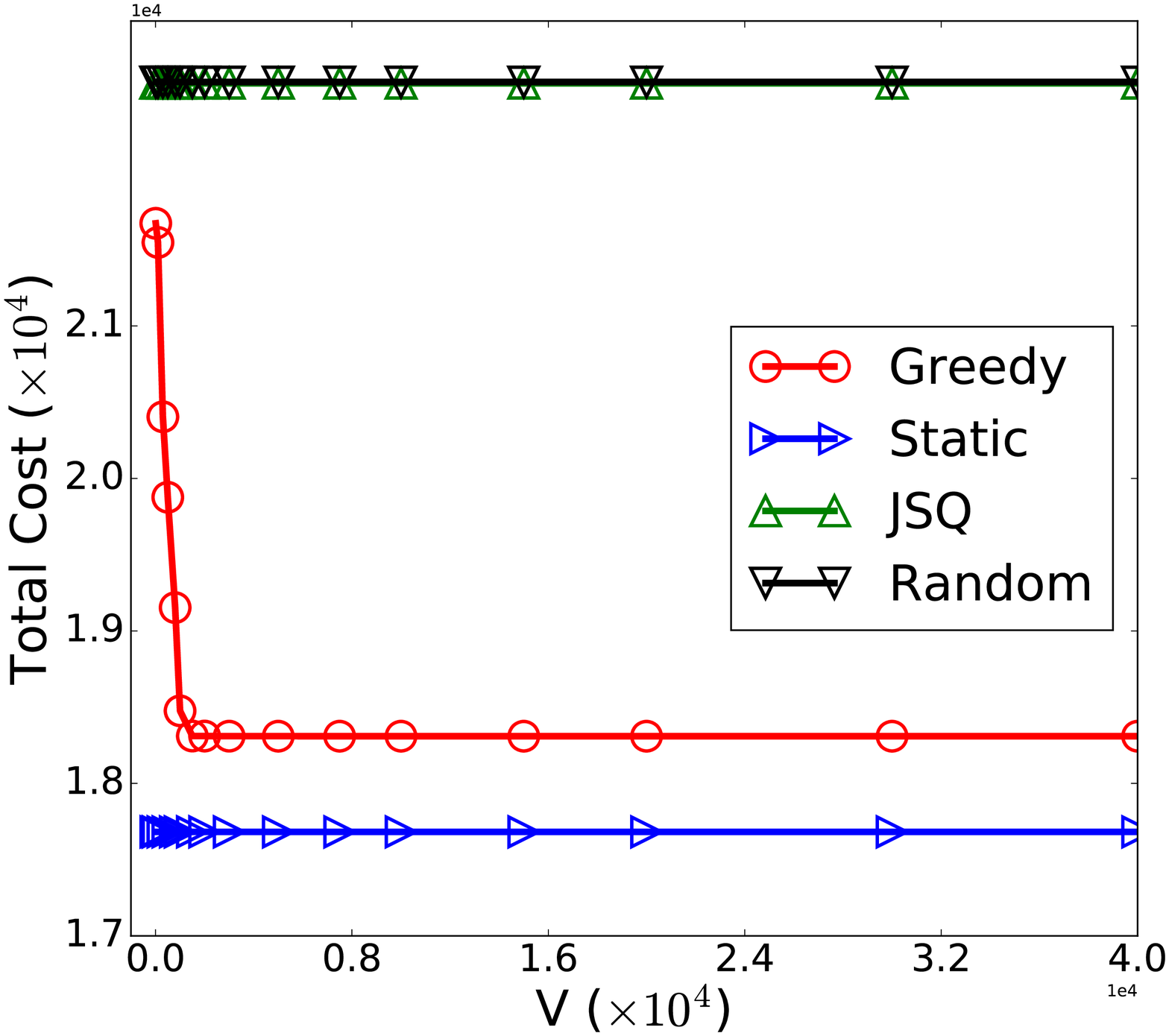}
 }
 \subfigure[Fat-tree topology] {
 \includegraphics[width=0.44\columnwidth]{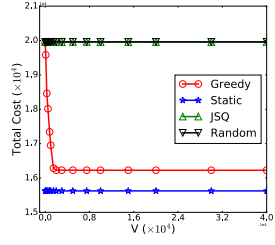}
 }
 \subfigure[Jellyfish topology] {
 \includegraphics[width=0.44\columnwidth]{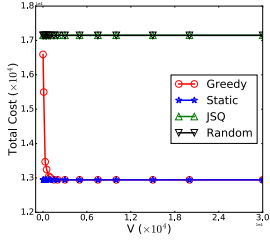}
 }
 \subfigure[F10 topology] {
 \includegraphics[width=0.44\columnwidth]{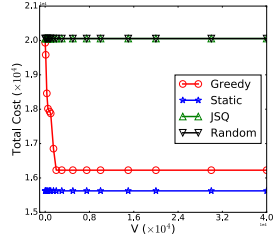}
 } 
 \caption{Communication cost comparison among four scheduling schemes under Canonical 3-Tiered, Fat-tree, Jellyfish, and F10 topology, respectively.}
 \label{cost_trace}
\end{figure}

Fig. \ref{cost_trace} presents a comparison among \textbf{Static}, \textbf{Random}, \textbf{JSQ}, and \textbf{Greedy} in terms of communication cost under those four topologies, respectively. We have the following observations. 

First, the communication cost under \textbf{Static} is the minimum among all schemes, which is consistent with its only goal of minimizing the overall communication cost. \textbf{Greedy} cuts down the communication cost with increasing $V$. Eventually, when $V$ is sufficiently large (around $1.0\times{10^{4}}$ to $2.0\times{10^4}$), communication cost stops decreasing and remains unchanged. Both \textbf{Random} and \textbf{JSQ} exhibit much higher communication costs, compared to \textbf{Greedy} and \textbf{Static}. This is due to the blindness of \textbf{Random} and \textbf{JSQ} to the communication cost to take when making scheduling decisions. 

\begin{figure}[!t]
\centering
 \subfigure[Canonical 3-Tiered topology] {
 \includegraphics[width=0.44\columnwidth]{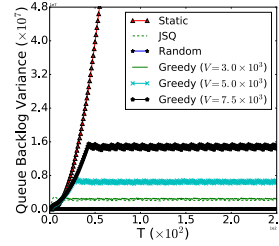}
 }
 \subfigure[Fat-tree topology] {
 \includegraphics[width=0.44\columnwidth]{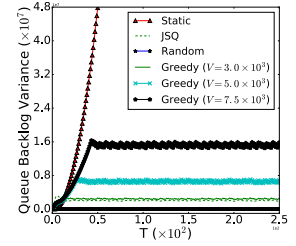}
}
\subfigure[Jellyfish topology] {
 \includegraphics[width=0.44\columnwidth]{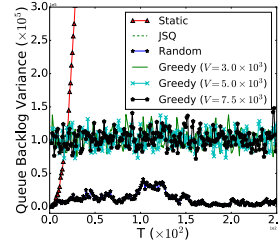}
 }
 \subfigure[F10 topology] {
 \includegraphics[width=0.44\columnwidth]{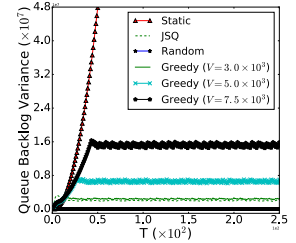}
}
 \caption{Variance of queue backlog size comparison among four scheduling schemes under Canonical 3-Tiered, Fat-tree, Jellyfish, and F10 topology, respectively.}
 \label{var_trace}
\end{figure}

Besides, we also observe that: there is still a gap between the communication cost of \textbf{Static} and the minimum cost that \textbf{Greedy} can reach. Here is the reason behind. With the growth of $V$'s value, \textbf{Greedy}'s scheduling behavior becomes increasingly similar to \textbf{Static}'s, which will lead to the reduction in cost and rise in queue backlogs. However, when the controllers' queue backlog size exceeds some threshold (about $2\times{V}$ in our simulation), especially for those close to hot spots, the scheduling decisions by \textbf{Greedy} and \textbf{Static} would be different again. For \textbf{Static}, its decision would continue pursuing minimum communication cost. This would accumulate even more requests onto heavily loaded controllers. For \textbf{Greedy}, however, some switches would rather turn to controllers with higher cost, so as to avoid the long queueing delay on those with lower cost. The difference in scheduling decisions would continue until the queue backlog size falls below the threshold again. Thus we can regard the gap as the cost that \textbf{Greedy} takes to stabilize the controllers' queue backlogs. The gap is much less significant in Jellyfish topology, because there are more switches (around $75\%$ in Jellyfish, higher than others) with multiple choices of minimum-cost controllers in Jellyfish than other topologies. Consequently, the range of request arrival fluctuation around the threshold (which results in different scheduling decisions of \textbf{Greedy} and \textbf{Static}) would be smaller, leading to a smaller gap.

Fig. \ref{var_trace} presents a comparison among the four schemes in terms of the variance of queue backlog size under those four topologies, respectively. In fact, smaller queue backlog size variance indicates better capability of load balancing. The variance of \textbf{Static} grows exponentially with time, showing that \textbf{Static} is incompetent in load balancing. The reason is that \textbf{Static} greedily associates switches with their nearest controllers, ignoring different controllers' loads, especially those controllers close to hot spots. When it comes to \textbf{Random} and \textbf{JSQ}, the variance is significantly lower, which shows the two schemes' advantage in load balancing. 
As for \textbf{Greedy}, in deterministic topologies (Fat-tree, F10, and Canonical 3-Tiered), its variance is in between the other three: the variance increases at the beginning and then remains stable soon after only about hundreds of time slots. With larger $V$, \textbf{Greedy} exhibits higher variance of queue backlog size, \textit{i.e.}, the load of controllers is more imbalanced. In contrast, in the random topology, \textit{i.e.} Jellyfish, increased $V$ in \textbf{Greedy} seems to have insignificant impact on the variance of queue backlogs. The reason is that Jellyfish has both smaller variance and average of shortest path lengths between switches and controllers than other topologies. Even though hot spots exist, the arriving requests would be spread more evenly to controllers in Jellyfish topology.

\begin{figure}[!t]
\centering
 \subfigure[Canonical 3-Tiered topology] {
 \includegraphics[width=0.465\columnwidth]{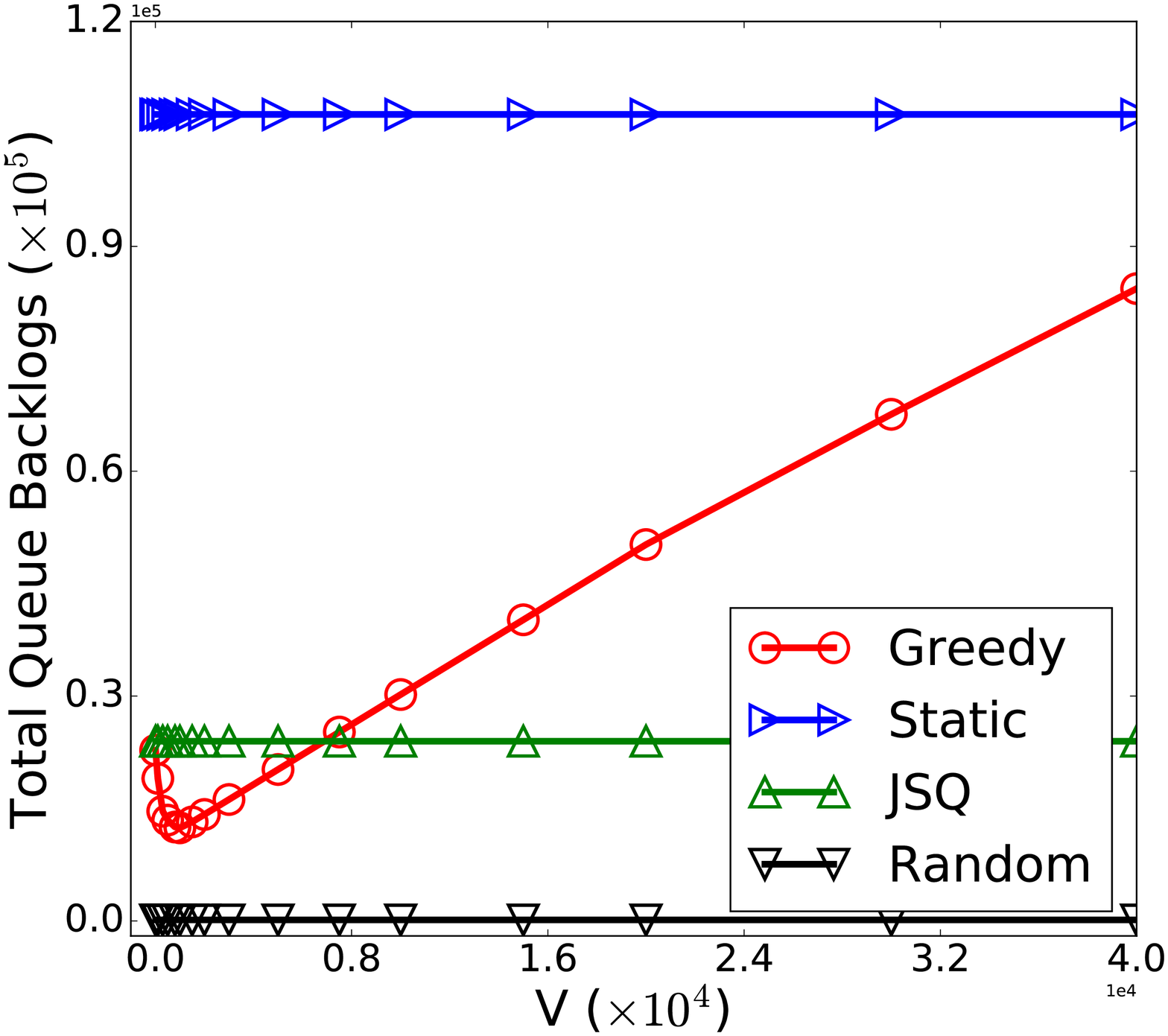}
 }
 \subfigure[Fat-tree topology] {
 \includegraphics[width=0.465\columnwidth]{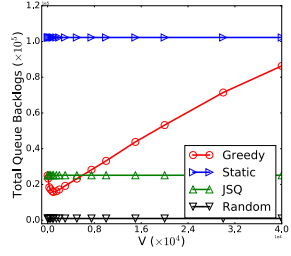}
 }
 \subfigure[Jellyfish topology] {
 \includegraphics[width=0.465\columnwidth]{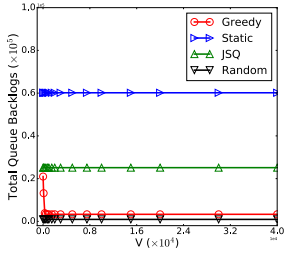}
 }
 \subfigure[F10 topology] {
 \includegraphics[width=0.465\columnwidth]{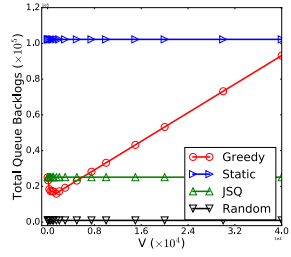}
 }
 \caption{Total queue backlog size comparison among four scheduling schemes under Canonical 3-Tiered, Fat-tree, Jellyfish, and F10 topology, respectively.}
 \label{qlen_trace}
\end{figure}

Fig. \ref{qlen_trace} shows a comparison among the four schemes in terms of the total queue backlog size under those four topologies, respectively.
The curves of \textbf{Static}, \textbf{JSQ}, and \textbf{Random} in Fig. \ref{qlen_trace} are very consistent with our observation from Fig. \ref{var_trace}. Intuitively, fixed the service rate on each controller, the more balanced the loads on control plane are, the more controllers' service are utilized, and hence the smaller of the total queue backlog size. In Fig. \ref{var_trace}, the variance of \textbf{Static} is high while that of \textbf{Random} and \textbf{JSQ} are much lower, so the total queue backlog size of \textbf{Static} is large while that of \textbf{Random} and \textbf{JSQ} is small in Fig. \ref{qlen_trace}. 
When it comes to \textbf{Greedy}, for deterministic topologies, we observe a declining trend at the very beginning, then the curve of total queue backlog size rises linearly after reaching a valley at around $1.0 \times 10^{3}$. 
The explanation is as follows. 

Consider the process in one time slot.
When $V$ is small, switches prefer controllers with shorter queues.\footnote{Note that \textbf{JSQ} is just a special case of \textbf{Greedy} with $V=0$.} A switch's scheduling decision is independent of the others'. This will lead to arriving requests being intensively uploaded to just few controllers. In this way, controllers close to hot spots are more likely to get heavier loads, though \textbf{Greedy} would adjust the load spread in the next time slot. 

As $V$ becomes larger, some switches would reach a tipping point and choose other controllers instead. As a result, this would mitigate the skewness of controllers' loads; \textit{i.e.}, the loads at control plane would become more and more balanced. This explains the declination of the curve. With the continual increasing in $V$, switches' interest in minimizing communication cost becomes dominant. Hence, the skewness of controllers' loads is aggravated and turns to linear rise.

\begin{figure}[!t]
\centering
 \subfigure[Poisson] {
 \includegraphics[width=0.44\columnwidth]{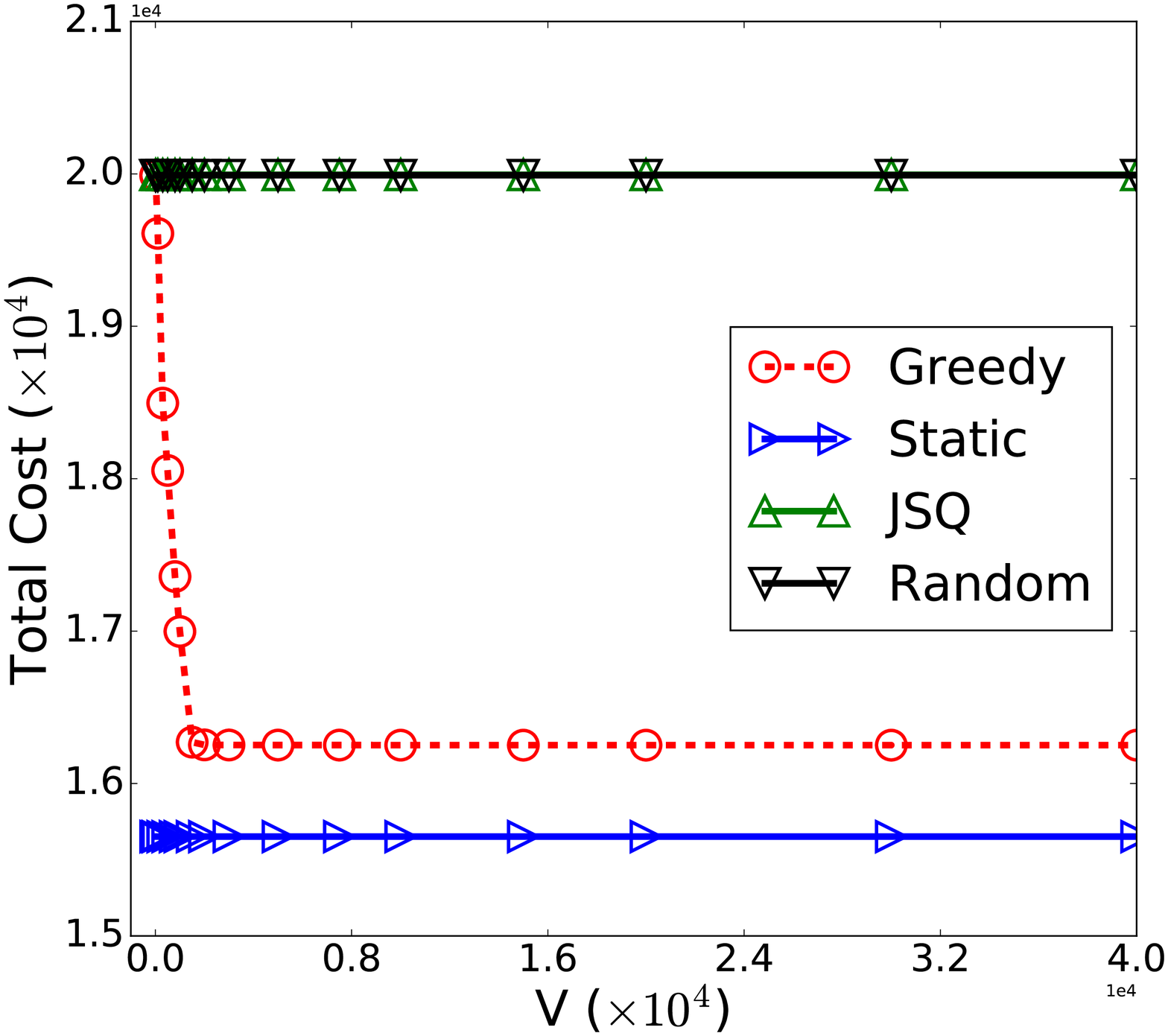}
 }
 \subfigure[Pareto] {
 \includegraphics[width=0.44\columnwidth]{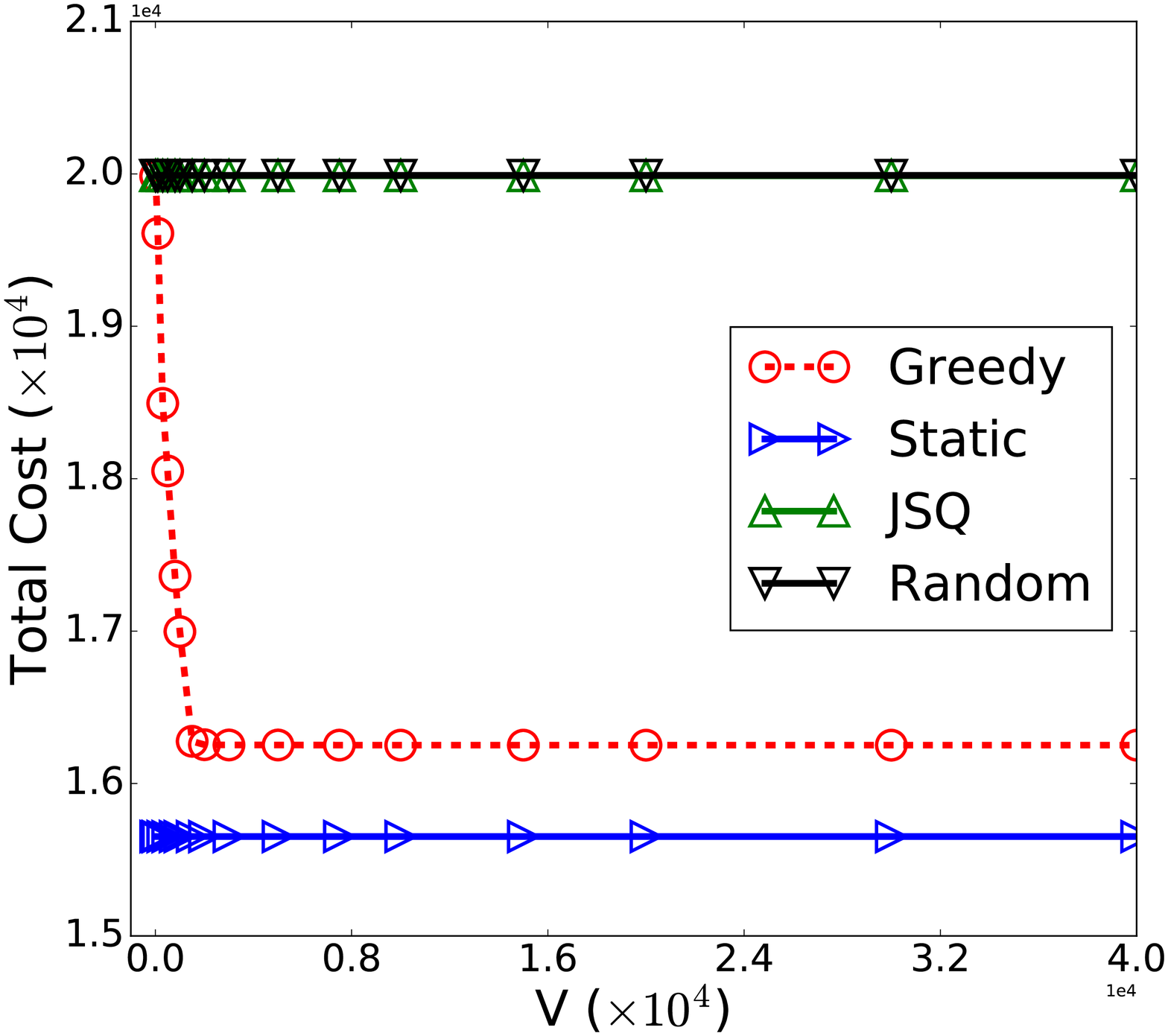}
 }
 \caption{Communication cost comparison among four scheduling schemes under Fat-tree topology, when the flow arrival follows Poisson and Pareto, respectively.}
 \label{cost}
\end{figure}

\begin{figure}[!t]
\centering
 \subfigure[Poisson] {
 \includegraphics[width=0.44\columnwidth]{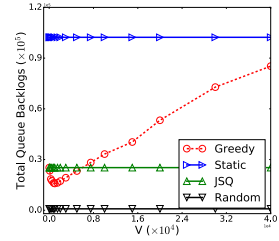}
 }
 \subfigure[Pareto] {
 \includegraphics[width=0.44\columnwidth]{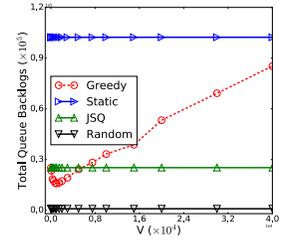}
 }
 \caption{Total queue backlog size comparison among four scheduling schemes under Fat-tree topology, when the flow arrival follows Poisson and Pareto, respectively.}
 \label{qlen}
\end{figure} 

When \textbf{Greedy} is applied in Jellyfish, its curve is very different from other three topologies. The curve decreases at the beginning and then stays at a low level constantly. To explain the difference, we notice that the variation of queue backlog size is highly related to the variance of request arrivals among controllers. To measure the variance, given a controller $j$, we define $j$'s \emph{minimum-cost request arrival rate} as the summation of the arrival rates from all switches to whom $j$ is one of those controllers with minimum cost. Thereby greater variance of controllers' minimum-cost request arrival rates will result in greater skewness of controllers' loads, when switches put less concern on queue backlogs and more on minimizing communication cost. 
In our setting, the variance of request arrivals among controllers is $1.62\times{10^{5}}$ in Jellyfish, while for Fat-tree, F10, and Canonical 3-Tiered, the values are $7.43\times{10^{5}}$, $7.43\times{10^{5}}$, and $5.25\times{10^{5}}$, respectively. Consequently, for topologies (\textit{e.g.} Fat-tree, F10, and Canonical 3-Tiered) with more imbalanced controllers' minimum-cost request arrival rates, the skewness of controllers' loads turns significant again, which results in the linear rising curve. 
For Jellyfish, however, because incoming requests are spread more evenly among controllers, increased $V$ has insignificant impact on the skewness of controllers' queue backlogs; hence its curve stays at a low level and is quite different from the others.

Note that both the curves under deterministic and random topologies are consistent with our theoretical analysis in (\ref{th1}), since (\ref{th1}) shows just the upper bound of total queue backlog size. The actual variation of queue backlog depends on the characteristics of underlying topologies. The more balanced switches and controllers are connected, the less significant queue backlog skewness there will be. 

In addition to trace-driven simulation, we also conduct the comparison with two kinds of flow arrival processes, \textit{i.e.}, \emph{Poisson} and \emph{Pareto} processes. They two are widely adopted in traffic analysis. For Poisson process, we set its arrival rate as $5.88$; while for Pareto process, we set its shape parameter as $2$ and its scale parameter as $2.94$.
We only show the simulation results under Fat-tree topology, because the simulation results in other three topologies are qualitatively similar.  Fig. \ref{cost} shows the communication cost comparison when the flow arrival process follows Poisson and Pareto, respectively. Fig. \ref{qlen} shows the total queue backlog size comparison when the flow arrival process follows Poisson and Pareto processes, respectively. We can see from these figures that the scheduling policies perform qualitatively consistent under different arrival processes.

In summary, among four schemes, \textbf{Static} is on the one end of performance spectrum: it minimizes communication cost while incurring extremely large queue backlogs; both \textbf{Random} and \textbf{JSQ} are on the other end of performance spectrum: they maintain the total queue backlog at a low level while incurring much larger communication costs. In contrast, our \textbf{Greedy} scheme achieves a trade-off between minimization of communication costs and minimization of queue backlogs. Through a tunable parameter $V$, we can achieve different degrees of balance between cost minimization and latency (queue backlog) minimization.

\section{Conclusion}

In this paper, we studied the joint optimization problem of dynamic switch-controller association and dynamic control devolution for SDN networks. We formulated the problem as a stochastic network optimization problem, aiming at minimizing the long-term average summation of total communication cost and computational cost while maintaining low time-average queue backlogs. We proposed an efficient online greedy algorithm, which yields a long-term average sum of communication cost and computational cost within $O(1 \slash V)$ of optimality, with a trade-off in an $O(V)$ queue backlog size for any positive control parameter $V$. Extensive simulation results show the effectiveness and optimality of our online algorithm, and the ability to maintain a tunable trade-off compared to other dynamic association schemes. 


\newpage
\appendices

\section{Problem Transformation by Opportunistically Minimizing an Expectation}

By minimizing the upper bound of the drift-plus-penalty expression (\ref{cond-v-drift}), the time average of communication cost can be minimized while stabilizing the network of request queues. 
We denote the objective function of (\ref{opt-pr}) at time slot $t$ by $J_t(\mathbf{X})$ and its optimal solution by $\mathbf{X}^* \in \mathcal{A}$.

Therefore, for any other scheduling decision $\mathbf{X} \in \mathcal{A}$ made during time slot $t$, we have
\begin{equation}
	\begin{array}{c}
		J_t(\mathbf{X}) \ge J_t(\mathbf{X}^*)
	\end{array}
\end{equation}
Taking the conditional expectation on both sides conditional on $\mathbf{Q}^c(t)$, we have
\begin{equation}
	\begin{array}{c}
		E\left[J_t(\mathbf{X})\,|\,\mathbf{Q}^c(t)\right] \ge E\left[J_t(\mathbf{X}^*)\,|\,\mathbf{Q}^c(t)\right]
	\end{array}
\end{equation}
for any $\mathbf{X} \in \mathcal{A}$. In such a way, instead of directly solving the long-term stochastic optimization problem (\ref{ori-opt-pr}), we can opportunistically choose a feasible association to solve problem (\ref{opt-pr}) during each time slot. 

\section{Proof of Theorem 1}
Given an association $\mathbf{X} \in \mathcal{A}$, for switch $i \in \mathcal{S}$, we define
\begin{equation}\label{yi-def}
\begin{array}{c}
	\mathbf{Y}_i \triangleq 1 - \sum_{j \in \mathcal{C}} \mathbf{X}_{i,j}
\end{array}
\end{equation}

Then with $L(\mathbf{Q}(t))$ defined in (\ref{lyqueue}), we have 
\begin{equation}\label{lyap-diff}
	\begin{array}{cl}
		& L(\mathbf{Q}(t+1)) - L(\mathbf{Q}(t)) \\
		= & \displaystyle \frac{1}{2} \left( \sum_{j \in \mathcal{C}} \left[ \left(Q^c_j(t+1)\right)^2 - \left(Q^c_j(t)\right)^2 \right] + \sum_{i \in \mathcal{S}} \left[ \left(Q^s_i(t+1)\right)^2 - \right. \right. \\
		& \displaystyle \left. \left. \left(Q^s_i(t)\right)^2 \right] \right) \\
		\le & \displaystyle \frac{1}{2} \sum_{j \in \mathcal{C}} \left\{ \left( Q^c_j(t) - B_j(t) + \sum_{i \in \mathcal{S}} \mathbf{X}_{i,j} \cdot {A}_{i}(t) \right)^2 - \left(Q^c_j(t)\right)^2 \right\} + \\
		& \displaystyle \frac{1}{2} \sum_{i \in \mathcal{S}} \left\{ \left( Q^s_i(t) - U_i(t) + \mathbf{Y}_i \cdot {A}_{i}(t) \right)^2 - \left(Q^s_i(t)\right)^2 \right\} \\
		= & \displaystyle \frac{1}{2} \sum_{j \in \mathcal{C}} \left\{ 2Q^c_j(t) \cdot \left( \sum_{i \in \mathcal{S}} \mathbf{X}_{i,j}\cdot {A}_{i}(t) - B_j(t) \right) + \right. \\
		& \displaystyle \left. \left( \sum_{i \in \mathcal{S}} \mathbf{X}_{i,j} \cdot {A}_{i}(t) - B_j(t) \right)^2 \right\} + 
		\frac{1}{2} \sum_{i \in \mathcal{S}} \displaystyle \left\{ \left( \mathbf{Y}_i \cdot {A}_{i}(t) - U_i(t) \right)^2 \right. \\
		& \displaystyle \left. + 2Q^s_i(t) \cdot \left( \mathbf{Y}_i \cdot {A}_{i}(t) - U_i(t) \right) \right\}
		\\
		\le & \displaystyle \sum_{j \in \mathcal{C}} \left\{ Q^c_j(t) \cdot \left( \sum_{i \in \mathcal{S}} \mathbf{X}_{i,j} \cdot {A}_{i}(t) - B_j(t) \right) + \right. \\
		& \displaystyle \left. \frac{ (\sum_{i \in \mathcal{S}} \mathbf{X}_{i,j}\cdot {A}_{i}(t))^2 + (B_j(t))^2}{2} \right\} + \\
		& \displaystyle \sum_{i \in \mathcal{S}} \left\{ Q^s_i(t) \cdot \left( \mathbf{Y}_{i} \cdot {A}_{i}(t) - U_i(t) \right) +
		\frac{ (Y_i \cdot {A}_{i}(t))^2 + (U_i(t))^2}{2} \right\} \\
	\end{array}	
\end{equation}

Then with the definition of $\Delta(\mathbf{Q}(t))$ in (\ref{cond-drift}), we have
\begin{equation}\label{lyap-delta1}
	\begin{array}{cl}
		& \Delta(\mathbf{Q}(t)) \\
		= & E\left\{ L\left(\mathbf{Q}(t+1)\right) - L\left( \mathbf{Q}(t) \right)\,|\,\mathbf{Q}(t) \right\} \\
		\le & \displaystyle E\left\{ \sum_{j \in \mathcal{C}} Q^c_j(t) \cdot \left( 
		\sum_{i \in \mathcal{S}} \mathbf{X}_{i,j}(t) A_i(t) - B_j(t)
		\right) \,|\, \mathbf{Q}(t) \right\} + \\
		  & \displaystyle E\left\{ \sum_{i \in \mathcal{S}} Q^s_i(t) \cdot \left( 
		\mathbf{Y}_{i}(t) A_i(t) - U_i(t)
		\right) \,|\, \mathbf{Q}(t) \right\} + \\
		& \displaystyle \frac{1}{2} E \left\{ \sum_{j \in \mathcal{C}} \left[ \left( \sum_{i \in \mathcal{S}} \mathbf{X}_{i,j} A_i(t) \right)^2  + \left( B_j(t) \right)^2 \right] \,|\,\mathbf{Q}(t) \right\} + \\
		& \displaystyle \frac{1}{2} E \left\{  \sum_{i \in \mathcal{S}} \left[ \left( \mathbf{Y}_{i} A_i(t) \right)^2  + \left( U_i(t) \right)^2 \right] \,|\,\mathbf{Q}(t) \right\} \\
		= & \displaystyle \sum_{j \in \mathcal{C}} Q^c_j(t) \cdot E \left\{ \left( 
		\sum_{i \in \mathcal{S}} \mathbf{X}_{i,j}(t) A_i(t) - B_j(t)
		\right) \,|\, \mathbf{Q}(t) \right\} + \\
		  & \displaystyle \sum_{i \in \mathcal{S}} Q^s_i(t) \cdot E \left\{ \left( 
		\mathbf{Y}_{i}(t) A_i(t) - U_i(t)
		\right) \,|\, \mathbf{Q}(t) \right\} + \\
		& \displaystyle \frac{1}{2} \sum_{j \in \mathcal{C}} \left[ \left( \sum_{i \in \mathcal{S}} \mathbf{X}_{i,j} A_i(t) \right)^2  + \left( B_j(t) \right)^2 \right] + \\
		& \displaystyle \frac{1}{2} \sum_{i \in \mathcal{S}} \left[ \left( \mathbf{Y}_{i} A_i(t) \right)^2  + \left( U_i(t) \right)^2 \right] \\
	\end{array}
\end{equation}
The last equality in (\ref{lyap-delta1}) holds because of conditional expectation on $\mathbf{Q}(t)$, then both $Q_i^s(t)$ and $Q_j^c(t)$ can be regarded as a constant. Besides, the queueing process $\{\mathbf{Q}(t)\}$ is independent of the arrival process $\{\mathbf{A}(t)\}$ and service process $\{\mathbf{B}(t)\}$, $\{\mathbf{U}(t)\}$. Hence, the last two terms are independent of $\mathbf{Q}(t)$. 

Now consider the last two terms in (\ref{lyap-delta1}). We have
\\
\begin{equation}\label{lyap-delta2}
	\begin{array}{cl}
		& \displaystyle \frac{1}{2} \sum_{j \in \mathcal{C}} \left[ \left( \sum_{i \in \mathcal{S}} \mathbf{X}_{i,j} A_i(t) \right)^2  + \left( B_j(t) \right)^2 \right] + \\
		& \displaystyle \frac{1}{2} \sum_{i \in \mathcal{S}} \left[ \left( \mathbf{Y}_{i} A_i(t) \right)^2  + \left( U_i(t) \right)^2 \right] \\
		= & \displaystyle \frac{1}{2} \left[ \sum_{j \in \mathcal{C}} \left( B_j(t) \right)^2 + \sum_{i \in \mathcal{S}} \left( U_i(t) \right)^2 \right] + \\
		& \displaystyle \frac{1}{2} \sum_{j \in \mathcal{C}} \left[ \left( \sum_{i \in \mathcal{S}} \mathbf{X}_{i,j} A_i(t) \right)^2 \right] +  \\
		& \displaystyle \frac{1}{2} \sum_{i \in \mathcal{S}} \left[ \left( (1 - \sum_{j \in \mathcal{C}} \mathbf{X}_{i,j}) \cdot A_i(t) \right)^2 \right] \\
	\end{array}
\end{equation}
Then by taking expectation on (\ref{lyap-delta2}), the following holds
	\begin{align*}
		& \displaystyle E\left\{ \frac{1}{2} \sum_{j \in \mathcal{C}} \left[ \left( \sum_{i \in \mathcal{S}} \mathbf{X}_{i,j} A_i(t) \right)^2  + \left( B_j(t) \right)^2 \right] + \right. \\
		& \displaystyle \left. \frac{1}{2} \sum_{i \in \mathcal{S}} \left[ \left( \mathbf{Y}_{i} A_i(t) \right)^2  + \left( U_i(t) \right)^2 \right] \right\} \\
		= & \displaystyle E \left\{ \frac{1}{2} \left[ \sum_{j \in \mathcal{C}} \left( B_j(t) \right)^2 + \sum_{i \in \mathcal{S}} \left( U_i(t) \right)^2 \right] + \right. \\
		& \displaystyle \frac{1}{2} \sum_{j \in \mathcal{C}} \left[ \left( \sum_{i \in \mathcal{S}} \mathbf{X}_{i,j} A_i(t) \right)^2 \right] +  \\
		& \displaystyle \left. \frac{1}{2} \sum_{i \in \mathcal{S}} \left[ \left( (1 - \sum_{j \in \mathcal{C}} \mathbf{X}_{i,j}) \cdot A_i(t) \right)^2 \right] \right\} \\
		= & \displaystyle \frac{1}{2} \left[ \sum_{j \in \mathcal{C}} E\left\{ \left( B_j(t) \right)^2 \right\} + \sum_{i \in \mathcal{S}} E\left\{ \left( U_i(t) \right)^2 \right\} \right] + \\
		& \displaystyle \frac{1}{2} \sum_{j \in \mathcal{C}} E \left\{ \left( \sum_{i \in \mathcal{S}} \mathbf{X}_{i,j} A_i(t) \right)^2 \right\} + \\
		& \displaystyle \left. \frac{1}{2} \sum_{i \in \mathcal{S}} E \left\{ (1 - \sum_{j \in \mathcal{C}} \mathbf{X}_{i,j})^2 \cdot \left(A_i(t)\right)^2 \right\} \right\} \\
		= & \displaystyle \frac{1}{2} \left[ \sum_{j \in \mathcal{C}} E\left\{ \left( B_j(t) \right)^2 \right\} + \sum_{i \in \mathcal{S}} E\left\{ \left( U_i(t) \right)^2 \right\} \right] + \\
		& \displaystyle \frac{1}{2} \sum_{j \in \mathcal{C}} E \left\{ \sum_{i \in \mathcal{S}} \mathbf{X}_{i,j}^2 \left(A_i(t)\right)^2 + 2\sum_{i < i'} \mathbf{X}_{i,j}\mathbf{X}_{i',j} A_{i}(t) A_{i'}(t) \right\} + \\
		& \displaystyle \frac{1}{2} \sum_{i \in \mathcal{S}} E \left\{ (1 - \sum_{j \in \mathcal{C}} \mathbf{X}_{i,j})^2 \cdot \left(A_i(t)\right)^2 \right\} \numberthis \\
	\end{align*}
Remind that the request arrival processes $\{\mathbf{A}(t)\}$ are independent and they are also independent of $\mathbf{X}_{i,j}$ for $(i,j) \in \mathcal{S} \times \mathcal{C}$. Then we have
	\begin{align*}
		& \displaystyle E\left\{ \frac{1}{2} \sum_{j \in \mathcal{C}} \left[ \left( \sum_{i \in \mathcal{S}} \mathbf{X}_{i,j} A_i(t) \right)^2  + \left( B_j(t) \right)^2 \right] + \right. \\
		& \displaystyle \left. \frac{1}{2} \sum_{i \in \mathcal{S}} \left[ \left( \mathbf{Y}_{i} A_i(t) \right)^2  + \left( U_i(t) \right)^2 \right] \right\} \\
		= & \displaystyle \frac{1}{2} \left[ \sum_{j \in \mathcal{C}} E\left\{ \left( B_j(t) \right)^2 \right\} + \sum_{i \in \mathcal{S}} E\left\{ \left( U_i(t) \right)^2 \right\} \right] + \\
		& \displaystyle \frac{1}{2} \left[ \sum_{j \in \mathcal{C}} \sum_{i \in \mathcal{S}} E \left\{ \mathbf{X}_{i,j}^2 \right\} E\left\{ \left(A_i(t)\right)^2 \right\} \right. + \\
		& \displaystyle \left. 2\sum_{i < i'} E\left\{ \mathbf{X}_{i,j} \right\} E\left\{ \mathbf{X}_{i',j} \right\} E\left\{ A_{i}(t) \right\} E\left\{ A_{i'}(t) \right\} \right] + \\
		& \displaystyle \frac{1}{2} \sum_{i \in \mathcal{S}} E \left\{ (1 - \sum_{j \in \mathcal{C}} \mathbf{X}_{i,j})^2 \right\} \cdot E\left\{ \left(A_i(t)\right)^2 \right\} \\
		\le & \displaystyle \frac{1}{2} \left( C \cdot \max_{j \in \mathcal{C}} \{ E(B_j^2(t))\} + S \cdot \max_{i \in \mathcal{S}} \{ E(U_i^2(t))\} + \right. \\
		& \displaystyle \max_{i \in \mathcal{S}} \{ E(A_i^2(t))\} \left[ 
			\sum_{j \in \mathcal{C}} E \left\{ \left( \sum_{i \in \mathcal{S}} \mathbf{X}_{i,j} \right)^2 \right\} + \right. \\
		& \displaystyle \left. \left. \sum_{i \in \mathcal{S}} E\left\{ \left(1 - \sum_{j \in \mathcal{C}} \mathbf{X}_{i,j}\right)^2 \right\} 
		\right] \right) \\
		\le & \displaystyle \frac{1}{2} \max_{i,j} (E(B_j^2(t)), E(U_i^2(t)), E(A_i^2(t))) \cdot \\ 
		& \displaystyle \left( C + S + \underset{\mathbf{X} \in \mathcal{A}}{\text{max}} \left\{ 
				\sum_{j \in \mathcal{C}} \left( \sum_{i \in \mathcal{S}} \mathbf{X}_{i,j} \right)^2 
				+ \sum_{i \in \mathcal{S}} (\mathbf{Y}_i(t))^2
			\right\} \right) \numberthis \label{final-ineq} \\
	\end{align*}
where the first inequality holds because of the following reasoning. We suppose $i^{*} \in \arg \max_{i \in \mathcal{S}} A_i(t)$. Then for any $i, i' \in \mathcal{S}$
\begin{equation}
	\begin{array}{cl}
		& \displaystyle E \left\{ A_{i}(t) \right\} \cdot E \left\{ A_{i'}(t) \right\} \le \displaystyle \left( E\left\{ A_{i^*}(t) \right\} \right)^2
	\end{array}
\end{equation}
As we know that $\text{Var}\left\{ A_{i^*}(t) \right\} \ge 0$, then 
\begin{equation}
	\begin{array}{cl}
		& E \left\{ A_{i}(t) \right\} \cdot E \left\{ A_{i'}(t) \right\} \le E \left\{ A^2_{i^*}(t) \right\} \\
	\end{array}
\end{equation}
Thus the first inequality in (\ref{final-ineq}) holds.  

Next, for $\mathbf{X} \in \mathcal{A}$, we focus on the upper bound of $\sum_{j \in \mathcal{C}} \left( \sum_{i \in \mathcal{S}} \mathbf{X}_{i,j} \right)^2 + \sum_{i \in \mathcal{S}} \left(1 - \sum_{j \in \mathcal{C}} \mathbf{X}_{i,j}\right)^2$.
At each time slot, $\mathbf{X}_{i,j} \in \left\{ 0, 1 \right\}$ and for each switch $i \in \mathcal{S}$, it must decide either to upload requests to one of controllers or process them locally. Then among all $\mathbf{X}_{i,j}$ (for all $(i,j) \in \mathcal{S} \times \mathcal{C}$) and $(1 - \sum_{j \in \mathcal{C}} \mathbf{X}_{i,j})$ (for $i \in \mathcal{S}$), there are exactly $|\mathcal{S}|$ of them that's equal to one. Let $a \in [0, |\mathcal{S}|]$ denote the number of switches that decide to upload requests to control plane, \textit{i.e.}, there are $a$ terms among all $\mathbf{X}_{i,j}$ (for $(i,j) \in \mathcal{S} \times \mathcal{C}$) that's equal to one.
Likewise, let $b \in [0, |\mathcal{S}|]$ denote the number of switches that process requests locally. Accordingly, we know that $a + b = |
\mathcal{S}|$. Besides, $\sum_{i \in \mathcal{S}} \left( 1 - \sum_{j \in \mathcal{C}} \mathbf{X}_{i,j} \right)^2 = b$ since there are exactly $b$ switches such that for any switch $i$ among them $\sum_{j \in \mathcal{C}} \mathbf{X}_{i,j} = 0$. 

Now we prove that the upper bound of $\sum_{j \in \mathcal{C}} \left( \sum_{i \in \mathcal{S}} \mathbf{X}_{i,j} \right)^2$ is $a^2$ and the bound is reached when all $a$ switches is associated with the same controller. We use $\mathcal{R}$ to denote the set of those $a$ switches. We introduce indicator $\mathcal{I}_{k,l}$ such that $\mathcal{I}_{k,l} = 1$ if switch $k$ and switch $l$ are associated with the same controller and $0$ otherwise. Therefore, for any switch-controller association $\mathcal{N} \subseteq \mathcal{R} \times \mathcal{C}$ such that $|\mathcal{N}| = a$, we have
\begin{equation}\label{to-cp-ineq}
	\begin{array}{cl}
		& \displaystyle \sum_{j \in \mathcal{C}} \left( \sum_{i \in \mathcal{S}} \mathbf{X}_{i,j} \right)^2 \\
		= & \displaystyle \sum_{j \in \mathcal{C}} \left\{ \sum_{i: (i,j) \in \mathcal{N}} \mathbf{X}_{i,j}^2 + 2 \cdot \sum_{\substack{i, i' \in \mathcal{R}: i < i' \\ \text{ and } \newline (i,j), (i',j) \in \mathcal{N}}} \mathbf{X}_{i,j} \mathbf{X}_{i',j} \right\} \\
		= & \displaystyle \sum_{(i,j) \in \mathcal{N}} \mathbf{X}_{i,j}^2 + 2 \cdot \sum_{j \in \mathcal{C}} \sum_{i, i': i < i'} \mathcal{I}_{i,i'} \mathbf{X}_{i,j} \mathbf{X}_{i',j} \\
		= & \displaystyle \sum_{(i,j) \in \mathcal{N}} \mathbf{X}_{i,j}^2 + 2 \cdot \sum_{i, i' \in \mathcal{R}: i < i'} \mathcal{I}_{i,i'} \\
		= & \displaystyle a + 2 \cdot \sum_{i, i' \in \mathcal{R}: i < i'} \mathcal{I}_{i,i'}
	\end{array}
\end{equation}
where the last equality holds because for any pair of switches $(i,i')$, $\mathcal{I}_{i,i'} = 1$ only when $i$ and $i'$ upload requests to the same controller. From (\ref{to-cp-ineq}), we know that the upper bound is reached when $\mathcal{I}_{i,i'} = 1$ for all $i, i' \in \mathcal{R}$, \textit{i.e.}, when all switches in $\mathcal{R}$ connected to the same switches. In such case, since there are $\frac{1}{2}a(a-1)$ pairs of different switches, then the upper bound of $\sum_{j \in \mathcal{C}} \left( \sum_{i \in \mathcal{S}} \mathbf{X}_{i,j} \right)^2$ is $a + a(a-1) = a^2$. Hence, 
\begin{equation}\label{tt-ineq}
	\begin{array}{cl}
		& \displaystyle \sum_{j \in \mathcal{C}} \left( \sum_{i \in \mathcal{S}} \mathbf{X}_{i,j} \right)^2 + \sum_{i \in \mathcal{S}} (1 - \sum_{j \in \mathcal{C}} \mathbf{X}_{i,j})^2 \\
		\le & \displaystyle a^2 + b \\
		= & a^2 + |\mathcal{S}| - a \\
		= & \left( a - \frac{1}{2} \right)^2 + |\mathcal{S}| - \frac{1}{4}
	\end{array}
\end{equation}
Now that $a$ is a non-negative integer and $0 \le a \le |\mathcal{S}|$, then the upper bound in (\ref{tt-ineq}) reaches its maximum value $|\mathcal{S}|^2$ when $a = |\mathcal{S}|$. In other words, the upper bound reaches maximum when all switches in $\mathcal{S}$ upload requests to the same controller. 
As a result, 
\begin{equation}\label{k-upper}
	\begin{array}{cl}
		& \displaystyle E \left\{ 
			\frac{1}{2} \sum_{j \in \mathcal{C}} \left[ \left( \sum_{i \in \mathcal{S}} \mathbf{X}_{i,j} A_i(t) \right)^2  + \left( B_j(t) \right)^2 \right] + \right. \\
		& \displaystyle \left. \frac{1}{2} \sum_{i \in \mathcal{S}} \left[ \left( \mathbf{Y}_{i} A_i(t) \right)^2  + \left( U_i(t) \right)^2 \right] \right\} \\
		\le & \displaystyle \frac{1}{2} \max_{i,j} (E(B_j^2(t)), E(U_i^2(t)), E(A_i^2(t))) \cdot \left( |\mathcal{C}| + |\mathcal{S}| + |\mathcal{S}|^2 \right) \\
		= & \displaystyle \frac{d_{max}}{2} \left( |\mathcal{C}| + |\mathcal{S}| + |\mathcal{S}|^2 \right) = K \\
	\end{array}
\end{equation}

We assume the whole control plane is capable of handling all requests from data plane in the mean sense. Therefore, 
for $j \in \mathcal{C}$, 
there exists $\epsilon^c_j > 0$ such that $E[B_j(t)-\sum_{i\in\mathcal{S}}\mathbf{X}_{i,j}A_i(t)\,|\,\mathbf{Q}^c(t)] = \epsilon^c_j$. Likewise, for $i \in \mathcal{S}$, 
there exists $\epsilon^s_i > 0$ such that $E[U_i(t)-\mathbf{Y}_i A_i(t)\,|\,\mathbf{Q}^c(t)] = \epsilon^s_i$.
Following (\ref{k-upper}) and the definition in (\ref{cond-v-drift}), after taking expectation on $\Delta_{V}( \mathbf{Q}(t) )$, we have
\begin{equation}
	\begin{array}{cl}
		& E \left\{ \Delta_{V}(\mathbf{Q}(t) \right\} \\
		\le & \displaystyle K + \sum_{j \in \mathcal{C}} E\left\{ Q_j^c(t) \right\} \cdot E\left\{ E\left\{ \sum_{i \in \mathcal{S}} \mathbf{X}_{i,j}(t) A_i(t) - B_j(t)\,|\,\mathbf{Q}(t)\right\} \right\} \\
		& \displaystyle + \sum_{i \in \mathcal{S}} E\left\{ Q_i^s(t) \right\} \cdot E\left\{ E\left\{ [1 - \sum_{j \in \mathcal{C}} \mathbf{X}_{i,j}(t)] A_i(t) - U_i(t)\,|\,\mathbf{Q}(t)\right\} \right\} \\
		& \displaystyle + V \cdot E\left\{ E\{ f(t) + g(t) |\mathbf{Q}(t) \} \right\} \\
		= & \displaystyle K + \sum_{j \in \mathcal{C}} E\left\{ Q_j^c(t) \right\} \cdot E\left\{ \sum_{i \in \mathcal{S}} \mathbf{X}_{i,j}(t) A_i(t) - B_j(t) \right\} \\
		& \displaystyle + \sum_{i \in \mathcal{S}} E\left\{ Q_i^s(t) \right\} \cdot E\left\{ [1 - \sum_{j \in \mathcal{C}} \mathbf{X}_{i,j}(t)] A_i(t) - U_i(t) \right\} \\
		& \displaystyle + V \cdot E \left\{ f(t) + g(t) \right\} \\
		\le & \displaystyle K - \epsilon^c \sum_{j \in \mathcal{C}} Q_j^c(t) - \epsilon^s \sum_{i \in \mathcal{S}} Q_i^s(t) + V \cdot (f^* + g^*) \\
	\end{array}
\end{equation}
where $\epsilon^c = \min_{j \in \mathcal{C}} \{ \epsilon^c_j \}$, $\epsilon^s = \min_{i \in \mathcal{S}} \{ \epsilon^s_i \}$. Expanding the term $E\left\{ \Delta_{V}(\mathbf{Q}(t)) \right\}$, then for any time slot $\tau$,  
\begin{equation}
	\begin{array}{cl}
		& \displaystyle E\left\{ L\left(\mathbf{Q}(\tau + 1) \right) - L\left(\mathbf{Q}(\tau) \right) \right\} + V \cdot E\left\{ f(\tau) + g(\tau) \right\} \\
		\le & \displaystyle K - \epsilon^c \sum_{j \in \mathcal{C}} E\left\{Q_j^c(\tau)\right\} - \epsilon^s \sum_{i \in \mathcal{S}} E\left\{Q_i^s(\tau)\right\} + V (f^* + g^*) \\
	\end{array}
\end{equation}
Summing over $\tau \in \{0,\,1,\,2,\,\dots,\,t-1\}$ for some $t > 0$, then
\begin{equation}
	\begin{array}{cl}
		& \displaystyle E \left\{ L\left(\mathbf{Q}(t) \right) - L\left(\mathbf{Q}(0) \right) \right\} + V  \cdot \sum_{\tau = 0}^{t-1} E\left[ f(\tau) + g(\tau) \right] \\
		\le & \displaystyle t \cdot K 
		- \epsilon^c \sum_{\tau = 0}^{t-1} \sum_{j \in \mathcal{C}} E\left\{Q_j^c(\tau)\right\} 
		- \epsilon^s \sum_{\tau = 0}^{t-1} \sum_{i \in \mathcal{S}} E\left\{Q_i^s(\tau)\right\} + \\
		& t \cdot V \cdot (f^* + g^*) \\		
	\end{array}
\end{equation}
By re-arrangement of terms at both sides and ignoring some non-negative term such as $E\left\{ L\left(\mathbf{Q}(t)\right) \right\}$ and $E\left\{ Q^c_j(t) \right\}$, with $\epsilon^c, \epsilon^s > 0$ and $V > 0$, we have
\begin{equation}\label{vfg}
	\begin{array}{cl}
		& \displaystyle V \cdot \sum_{\tau = 0}^{t-1} E\left[ f(\tau) + g(\tau) \right] \\
		\le & t \cdot V \cdot (f^* + g^*) + t \cdot K + E\left\{ L(\mathbf{Q}(0)) \right\} \\
	\end{array}
\end{equation}
\begin{equation}\label{ecq}
	\begin{array}{cl}
		& \displaystyle \epsilon^c \cdot \sum_{\tau = 0}^{t-1} \sum_{j \in \mathcal{C}} E\left\{Q_j^c(\tau)\right\} \\
		\le & t \cdot V \cdot (f^* + g^*) + t \cdot K + E\left\{ L(\mathbf{Q}(0)) \right\} \\
	\end{array}
\end{equation}
\begin{equation}\label{esq}
	\begin{array}{cl}
		& \displaystyle \epsilon^s \cdot \sum_{\tau = 0}^{t-1} \sum_{i \in \mathcal{S}} E\left\{Q_i^s(\tau)\right\} \\
		\le & t \cdot V \cdot (f^* + g^*) + t \cdot K + E\left\{ L(\mathbf{Q}(0)) \right\} \\
	\end{array}
\end{equation}

Then by dividing both sides of (\ref{vfg}) by $V \cdot t$, (\ref{ecq}) by $\epsilon^c \cdot t$, and (\ref{esq}) by $\epsilon^s \cdot t$, we have
\begin{equation}
	\begin{array}{cl}
		& \displaystyle \frac{1}{t} \cdot \sum_{\tau = 0}^{t-1} E\left[ f(\tau) + g(\tau) \right] \\
		\le & \displaystyle (f^* + g^*) + \frac{K}{V} + \frac{E\left\{ L(\mathbf{Q}(0)) \right\}}{V \cdot t} \\
	\end{array}
\end{equation}
\begin{equation}
	\begin{array}{cl}
		& \displaystyle \frac{1}{t} \sum_{\tau = 0}^{t-1} \sum_{j \in \mathcal{C}} E\left\{Q_j^c(\tau)\right\} \\
		\le & \displaystyle \frac{V \cdot (f^* + g^*) + K}{\epsilon^c} + \frac{E\left\{ L(\mathbf{Q}(0)) \right\}}{\epsilon^c \cdot t} \\
	\end{array}
\end{equation}
\begin{equation}
	\begin{array}{cl}
		& \displaystyle \frac{1}{t} \sum_{\tau = 0}^{t-1} \sum_{i \in \mathcal{S}} E\left\{Q_i^s(\tau)\right\} \\
		\le & \displaystyle \frac{V \cdot (f^* + g^*) + K}{\epsilon^s} + \frac{E\left\{ L(\mathbf{Q}(0)) \right\}}{\epsilon^s \cdot t} \\
	\end{array}
\end{equation}
At last, taking the limit as $t \to \infty$ for both equations, we have the desired results:
\begin{equation}
	\begin{array}{cl}
		& \displaystyle \lim_{t \to \infty} \frac{1}{t} \cdot \sum_{\tau = 0}^{t-1} E\left[ f(\tau) + g(\tau) \right] \le \displaystyle f^* + g^* + \frac{K}{V} \\
	\end{array}
\end{equation}
\begin{equation}
	\begin{array}{cl}
		& \displaystyle \lim_{t \to \infty} \frac{1}{t} \sum_{\tau = 0}^{t-1} \sum_{j \in \mathcal{C}} E\left\{Q_j^c(\tau)\right\} \le \frac{V \cdot (f^* + g^*) + K}{\epsilon^c} \\
	\end{array}
\end{equation}
\begin{equation}
	\begin{array}{cl}
		& \displaystyle \lim_{t \to \infty} \frac{1}{t} \sum_{\tau = 0}^{t-1} \sum_{i \in \mathcal{S}} E\left\{Q_i^s(\tau)\right\} \le \frac{V \cdot (f^* + g^*) + K}{\epsilon^s} \\
	\end{array}
\end{equation}

By setting $\epsilon = \frac{1}{2} \min\{ \epsilon^c, \epsilon^s \}$, the following desired result holds
\begin{equation}
	\begin{array}{rc}
		\displaystyle \underset{t \to \infty}{\lim \sup} \frac{1}{t} \sum_{\tau=0}^{t-1} \left[ \sum_{j \in \mathcal{C}} E\left\{ Q^c_j(\tau) \right\} + \sum_{i \in \mathcal{S}} E\left\{ Q^s_i(\tau) \right\} \right]
			& \le \\
			\displaystyle \frac{K + V \cdot (f^* + g^*) }{\epsilon} \\
	\end{array}
\end{equation}

\IEEEQED

\end{document}